\documentclass[twocolumn,showpacs,graphicx,floatfix,eqsecnum,aps,prb,nobibnotes,twocolumn,hyperref=pdftex,amsmath,longbibliography]{revtex4-1}
\usepackage{amsmath}
\usepackage{amssymb}
\usepackage{epsfig,color}
\usepackage{graphicx}
\usepackage{dcolumn}
\usepackage{bm}
\usepackage{multirow}
\usepackage{bbm}
\usepackage{enumerate}
\usepackage{graphicx}
\usepackage{verbatim}
\usepackage{hyperref}
\usepackage{float}
\hypersetup{colorlinks=true,linkcolor=blue,anchorcolor=blue,citecolor=blue,filecolor=blue,urlcolor=blue,bookmarksnumbered=true,pdfview=FitB}
\usepackage[utf8]{inputenc}
\usepackage[english]{babel}
\usepackage[maxfloats=256]{morefloats}

\newcommand{\btau}{\hat{\boldsymbol{\tau}}}
\newcommand{\ve}{\varepsilon}

\newcommand{\bk}{{\bf k}}

\newcommand{\bq}{{\bf q}}

\newcommand{\nn}{\nonumber}
\newcommand{\beq}{\begin{equation}}
\newcommand{\eeq}{\end{equation}}
\newcommand{\bea}{\begin{eqnarray}}
\newcommand{\eea}{\end{eqnarray}}
\newcommand{\bse}{\begin{subequations}}
\newcommand{\ese}{\end{subequations}}
\newcommand{\bwt}{\begin{widetext}}
\newcommand{\ewt}{\end{widetext}}

\newcommand{\bu}{{\bf u}}

\newcommand{\R}{\mathrm{Re}}

\newcommand{\bsu}{\begin{subequations}}
\newcommand{\esu}{\end{subequations}}
\newcommand{\bs}{\hat{\boldsymbol{\sigma}}}
\newcommand{\bS}{\hat{\boldsymbol{\Sigma}}}

\newcommand{\br}{{\bf r}}

\newcommand{\e}{\epsilon}
\newcommand{\eq}{\eqref}
\newcommand{\Eq}{Eq.~\eqref}

\newcommand{\vd}{v_\mathrm{D}}

\newcommand{\er}{\eqref}
\newcommand{\lr}{\lambda_\mathrm{R}}
\newcommand{\lz}{\lambda_\mathrm{VZ}}
\newcommand{\dz}{\Delta_\mathrm{Z}}
\newcommand{\dr}{\Delta_\mathrm{R}}
\newcommand{\dd}{\Delta_\mathrm{D}}
\newcommand{\vs}{\varsigma}

\newcommand{\bi}{\begin{itemize}}
\newcommand{\ei}{\end{itemize}}

\begin{document}
\title{
Collective spin modes in Fermi liquids with spin-orbit coupling}
\author{Dmitrii L. Maslov}
\email{maslov@ufl.edu}
\affiliation{Department of Physics, University of Florida, Gainesville, FL 32611-8440, USA}
\author{Abhishek Kumar}
\affiliation{Center for Materials Theory, Department of Physics and Astronomy, Rutgers University, Piscataway, NJ 08854, USA
} 
\author{Saurabh Maiti}
\affiliation{Department of Physics and Centre for Research in Molecular Modeling, Concordia University, Montreal, QC  H4B 1R6, Canada}
\date{\today}

\begin{abstract}
A combination of spin-orbit coupling and electron-electron interaction gives rise to a new type of collective spin modes, which correspond to oscillations of magnetization even in the absence of the external magnetic field. We review recent progress in theoretical understanding and experimental observation of such modes, focusing on three examples of real-life systems: a two-dimensional electron gas with Rashba and/or Dresselhaus spin-orbit coupling, graphene with proximity-induced spin-orbit coupling, and the Dirac state on the surface of a three-dimensional topological insulator.  This paper is dedicated to the 95th birthday of Professor Emmanuel I. Rashba.
\end{abstract}

\maketitle
\section{Introduction}
Spin-orbit coupling (SOC) plays an important and, sometimes decisive, role in many condensed matter systems, including two-dimensional (2D) electron and hole gases in semiconductor heterostructures,\cite{zutic:2004,winkler:book} non-centrosymmetric normal metals~\cite{samokhin:2009} and superconductors,\cite{sigrist:1991,mineev_sigrist} bismuth tellurohalides \cite{Bahramy:2011}, a variety of iridates and vanadates, \cite{balents:2014}, surface/edge states of three-dimensional (3D)/2D topological insulators,\cite{hasan:2010,hasan:2011,qi:2011,alicea:2012,moore:2012}
conducting states at oxide interfaces,\cite{cen:2009} 2D  transition metal dichalcogenides (TMD),\cite{manzeli:2017,heinz:2018} graphene on TMD substrates,\cite{morpurnat} atomic Bose\cite{lin:2009,lin:2011} and Fermi\cite{wang:2012, cheuk:2012}  gases in simulated non-Abelian magnetic fields, etc.
Coupling between electron spins and momenta leads to a number of fascinating consequences, such as the electric-dipole spin resonance (EDSR), \cite{schulte:2005,wilamowski:2007,wilamowski:2008} current-induced spin polarization, \cite{kato:2004,sih:2005,wunderlich:2005} persistent spin helices, \cite{schliemann:2003, bernevig:2006,koralek:2009} quantum spin \cite{bernevig:2006a,bernevig:2006b,koenig:2007}  and anomalous Hall effects,\cite{Chang:2013,Checkelsky:2014,Chang:2015} to name just a few. An interesting and still largely open question is the interplay between spin-orbit and electron-electron interactions. Such interplay gives rise to new phases of matter, e.g., topological Mott insulator,\cite{pesin:2010,balents:2013} gyrotropic and multipolar orders in normal metals,\cite{fu:2015} helical Fermi liquid (FL),\cite{lundgren:2015} Gor'kov-Rashba superconductor with mixed singlet-triplet order parameter,\cite{gorkov:2001}  topological Kondo insulators, \cite{dzero:2010} etc. It also affects in a non-trivial way many physical phenomena, e.g., optical conductivity,\cite{farid:2006,agarwal:2011} plasmon spectra,\cite{badalyan:2009,raghu:2010,maiti:2014} RKKY interaction,\cite{simon:2007,chesi:2010,badalyan:2010} non-analytic behavior of the spin susceptibility,\cite{zak:2010,zak:2012,miserev:2021} etc., and gives rise to spin-dependent electron-electron interaction.\cite{gindikin:2022}

In this paper, we review recent progress in theoretical understanding and experimental observation of a new type of collective spin modes in 2D FLs with SOC.  Such modes are perhaps the most direct manifestation of an interplay between spin-orbit and electron-electron interactions, as their existence hinges on both components being present. Unlike the conventional Silin mode in a partially spin-polarized FL,\cite{silin:1958} these modes exist even in the absence of an external magnetic field; in addition, they modify in a characteristic way the Silin mode if both SOC and magnetic field are present. As long as SOC is weak, the new modes correspond to oscillations of the magnetization which are decoupled from the oscillations of charge. The origin of the new modes can be traced to the effective spin-orbit magnetic field, which depends on the orientation and magnitude of the electron momentum, and also on the position of electron valley in the Brillouin zone (for multi-valley systems, such as graphene with proximity-induced SOC). Some of these modes have already been observed experimentally in Cd$_{1-x}$Mn$_{x}$Te quantum wells (in the presence of the magnetic field)\cite{perez:2007, baboux:2013,baboux:2015,perez:2016,perez:2017,perez:2019}  and in the surface state of a three-dimensional (3D) topological insulator (TI) Bi$_2$Se$_3$ (in zero magnetic field);\cite{kung:2017} however, many more predictions are still awaiting their experimental confirmation.

The rest of the paper is organized as follows. In Sec.~\ref{sec:singleparticle}, we introduce three systems considered in the rest of the article: 1) a 2D electron gas (2DEG) with Rashba and/or Dresselhaus SOC, 2) graphene with proximity-induced SOC, and 3) a Dirac helical state on the surface of a 3D TI. In Sec.~\ref{sec:FL}, we describe the single- and two-valley FL theories, which will be applied to study the collective spin modes in 2DEGs and graphene, respectively, given that SOC and/or magnetic field are weak. In Sec.~\ref{sec:noFL}, we explain why a FL theory cannot be applied to the cases of arbitrarily strong SOC and/or magnetic field. Sec.~\ref{sec:Silin} serves as a short reminder of collective modes in a FL without SOC, in general, and of the Silin modes in a partially spin-polarized FL, in particular. In Sec.~\ref{sec:2DEG}, we discuss collective spin modes in a 2DEG. Sec.~\ref{sec:RSOC} describes the FL theory for the case of Rashba SOC. In Sec.~\ref{sec:lattice}, we show that the FL kinetic equation for a 2DEG with Rashba and/or Dresselhaus SOC and in the presence of the magnetic field can be mapped onto an effective tight-binding model for an artificial one-dimensional (1D) lattice, whose sites are labeled by the projections of the angular momentum. Within this mapping, the role of FL interaction is to produce ``defects'', both of the on-site and bond types, and the collective modes arise as bound states due to such defects. In Sec.~\ref{sec:RSOCZ}, we illustrate how this mapping works for the case of a 2DEG with Rashba SOC and in the presence of the magnetic field using the $s$-wave approximation for the Landau function. Sec.~\ref{sec:Dirac} deals with collective spin modes in Dirac systems. In Sec.~\ref{sec:GR} we apply a two-valley version of the FL theory to graphene with proximity-induced SOC. In Sec.~\ref{sec:helical}, we derive the spectrum of inter-band spin excitations in a Dirac surface state within the ladder approximation. In Sec.~\ref{sec:space}, we discuss the spatial dispersion of collective spin modes. Sec.~\ref{sec:damping} is devoted to damping due to both disorder and electron-electron interaction. In Sec.~\ref{sec:exp}, we discuss both the current and future experiment. Sec.~\ref{sec:finiteq} summarizes the results of a series of Raman experiments on Cd$_{1-x}$Mn$_{x}$Te. In Sec.~\ref{sec:Bi2Se3}, we provide a summary of recent Raman spectroscopy of a collective spin mode on the surface of Bi$_2$Se$_3$. Sec.~\ref{sec:ESR_EDSR} contains the theoretical predictions for electron spin resonance (ESR) and EDSR experiments on graphene with proximity-induced SOC, both in zero and strong (compared to SOC) magnetic field. Our conclusions are given in Sec.~\ref{sec:concl}.

\section{Model single-particle Hamiltonians and electron-electron interaction}
\label{sec:models}
\subsection{Model single-particle Hamiltonians for spin-orbit coupling}
\label{sec:singleparticle}
Despite the variety of real-life systems, the effects of SOC on the electron spectrum can be described by just a few low-energy Hamiltonians, constructed by using the symmetry arguments.
In this paper, we will consider three examples of two-dimensional electron systems: a two-dimensional electron gas (2DEG) sandwiched between two dissimilar semiconductors, monolayer graphene with substrate-induced SOC, and the surface state of a three-dimensional (3D) topological insulator (TI).

 The effect of SOC on a 2DEG sandwiched between two dissimilar centrosymmetric semiconductors is described by the venerable Rashba Hamiltonian\cite{rashba:1959,bychkov:1984}
\bea
\hat H_{\text{R}}=\frac{k^2}{2m}+\alpha(\bs\times\bk)\cdot \hat {\bf z}=\frac{k^2}{2m}+\alpha\left(k_y\hat\sigma_x-k_x\hat\sigma_y\right),\nn\\
\label{HR}
\eea
where $\bs=(\hat\sigma_x,\hat\sigma_y,\hat\sigma_z)$ is a vector of Pauli matrices which describe electron spin, $\hat {\bf z}$ is a unit vector along the normal to the 2DEG plane, and $\alpha$ is a phenomenological parameter (with units of velocity). 
The simplest way to arrive at this Hamiltonian is to notice that a combination $(\bs\times\bk)\cdot \hat {\bf z}$ is the only scalar which can be formed out of an axial vector ($\bs$) and two polar vectors ($\bk$ and $\hat {\bf z}$).

In a bulk non-centrosymmetric semiconductor, e.g., of the A3B5 family (GaAs, CdTe, etc.), symmetry allows for a cubic (Dresselhaus) coupling between momentum and spin.\cite{dresselhaus:1955} A quantum well on the surface of such a semiconductor is described 
by a 2D Dresselhaus Hamiltonian with linear coupling between spin and momentum obtained by projecting the bulk Dresselhaus term onto the quantum-well plane.\cite{dyakonov:1986,dyakonov:book}  A particular form of the 2D Dresselhaus Hamiltonian depends
on the orientation of the quantum-well plane with respect to crystallographic axes.  We will consider the most common case of a quantum well grown along the (001) direction. With the $x$-axis along the (100) direction and $y$-axis along the (010) the direction, the Dresselhaus  Hamiltonian reads
\bea
\hat H_{\text{D}}=\frac{k^2}{2m}+\beta\left(k_x\hat\sigma_x-k_y\hat\sigma_y\right).\label{HD}
\eea
In heterostructures made from non-centrosymmetric semiconductors the Rashba and Dresselhaus types of SOC usually occur simultaneously, and the total spin-orbit part of the Hamiltonian is the sum of the Rashba and Dresselhaus terms.  Without loss of generality, we assume that $\alpha,\beta>0$.

Another popular system is graphene adsorbed on a transition-metal-dichalcogenide (TMD) substrate, such as WS$_2$, WSe$_2$ and MoS$_2$.\cite{Avsar:2014,morpurgo_2015,morpurgo_2016,Yang:2016,shi:2017,Dankert:2017,Ghiasi:2017,wakamura:2018,schonenberger:2018,Omar:2018,Benitez:2018,wakamura:2019,Wang:2019,Island:2019} 
In this case, the induced SOC is expected to be a mixture of two types:\cite{morpurgo_2015,cummings:2017,garcia:2018} of Rashba SOC
and 
 valley-Zeeman (VZ) or Ising  SOC, which acts as an out-of-plane magnetic field whose direction alternates between the $K$ and $K'$ valleys of graphene. We focus on the case of monolayer graphene with proximity-induced SOC,
described by the following low-energy Hamiltonian:
\bea
\label{ham0}
\hat{H}_{\text{GR}} &=& \vd(\tau_z \hat{\nu}_x k_x + \hat{\nu}_y k_y) + \Delta \hat\nu _z 
 +  \frac{\lambda_{\rm R}}{2} (\tau_z \hat{\sigma}_y \hat{\nu}_x - \hat{\sigma}_x \hat{\nu}_y)\nn\\
&&  + \frac{\lz}{2} \tau_z \hat{\sigma}_z,
 \eea
where $\vd$ is the Dirac velocity, $\textbf{k}$ is the electron momentum measured either from the $K$ or $K'$ point of the graphene Brillouin zone, $\lr$ and $\lz$ are the coupling constants of the Rashba  and VZ spin-orbit interactions, respectively, $\Delta$ is the gap due to substrate-induced asymmetry between the A and B sites of the honeycomb lattice, $\hat{\nu}_i$ ($i=x,y$) are the Pauli matrices in the sublattice (pseudospin) subspace, and $\tau_z=\pm 1$ labels the $K$ and $K'$ points. (An expression of the type $\hat\sigma_i\hat\nu_j$ is to be understood as a tensor product of two matrices. A single matrix in one of the subspaces implies that it is tensor product with a unity matrix in the other subspace.)  In \Eq{ham0}, we neglected the intrinsic, Kane-Mele type of SOC,\cite{KM} which is much weaker than the induced ones.\footnote{Theoretical estimates place the Kane-Mele coupling in the range from 1\,$\mu$eV \cite{min:2006,yao:2007} to 25-50\,$\mu$eV,\cite{trickey:2007,konschuh:2010} while
a recent ESR experiment reports the value of $42.2$\, $\mu$eV \cite{sichau:2019}} 

Finally, the surface of a 3D TI, e.g., Bi$_2$Se$_3$, harbors a Dirac helical state.
If the hexagonal warping of the energy contours due to underlying crystal lattice can be neglected, this state  is described by the Rashba-like Hamiltonian without the parabolic term\cite{fu:2009}
\bea
\hat H_{\text{TI}}= \vd(\bs\times\bk)\cdot\hat{\bf z}.\label{HTI}
\eea

In all cases presented above, the effect of an in-plane magnetic field of magnitude $B$ and at angle $\gamma$ to the $x$-axis is accounted for by adding the usual Zeeman term
\bea
\hat H_{\text{Z}}=\frac 12 \Delta_{\text{Z}}\left(\cos\gamma\;\hat\sigma_x+\sin\gamma\;\hat\sigma_y\right) \label{HZ}
\eea
to the corresponding Hamiltonian. 
Here, $\Delta_{\text{Z}}=g\mu_BB$ is the Zeeman splitting, $g$ is the effective Land{\'e} factor (assumed to be isotropic) and $\mu_B$ is the Bohr magneton.

The properties of Hamiltonians \eq{HR}-\eq{HTI} are well-understood by now and we will not reproduce the known results here. It suffices to say that Rashba and Dresselhaus types of SOC lead to spin textures in the momentum space, which can be interpreted as the effect of an effective magnetic field,
while the VZ type of SOC in \eq{ham0} acts as an out-of-plane magnetic field whose orientation alternates between the $K$ and $K'$ valleys.

\subsection{Models of electron-electron interaction}
\label{sec:FL}
Since the focus of this article is on the collective modes,
we need to invoke the electron-electron interaction as it is essential to induce collective behaviour. In all cases, we assume that our system is doped, such that the Fermi energy ($E_F$) is significantly larger than the spin-orbit and/or Zeeman splitting of the electron spectrum.  An exception is the surface state of a TI, where  SOC forms the spectrum rather than modifies the already existing one. With this exception, SOC can be treated as a perturbation imposed on a single-valley (2DEG) or two-valley (graphene) Fermi liquid.  Modulo renormalizations by the interaction, the spin-orbit and Zeeman energy scales determine the frequencies of the corresponding collective modes. Therefore, to leading order in these energy scales, one can neglect the effect of SOC and magnetic field on the Landau interaction function. For the single-valley case, the latter is given by the usual form \cite{agd:1963,lifshitz:1980,baym:book}
 \bea
 N_F^*\hat f(\bk,\bk')
 =
 F^s(\vartheta) +
\bs
 \cdot\bs'
 F^a(\vartheta),
 \label{FSU2}
 \eea
 where $N_F^*$ is the total (including the spin degeneracy), renormalized density of states at the Fermi energy, $\vartheta$ is the angle between the momenta $\bk$ and $\bk'$ of two quasiparticles on the Fermi surface ($|\bk|=|\bk'|=k_F$), functions $F^s$ and $F^a$ describe the interaction in the symmetric (direct) and asymmetric (exchange) channels, and unprimed/primed $\sigma$-matrices refer to the spin state of the first/second quasiparticle.
 
\begin{figure*}
\centering
\includegraphics[scale=0.45]{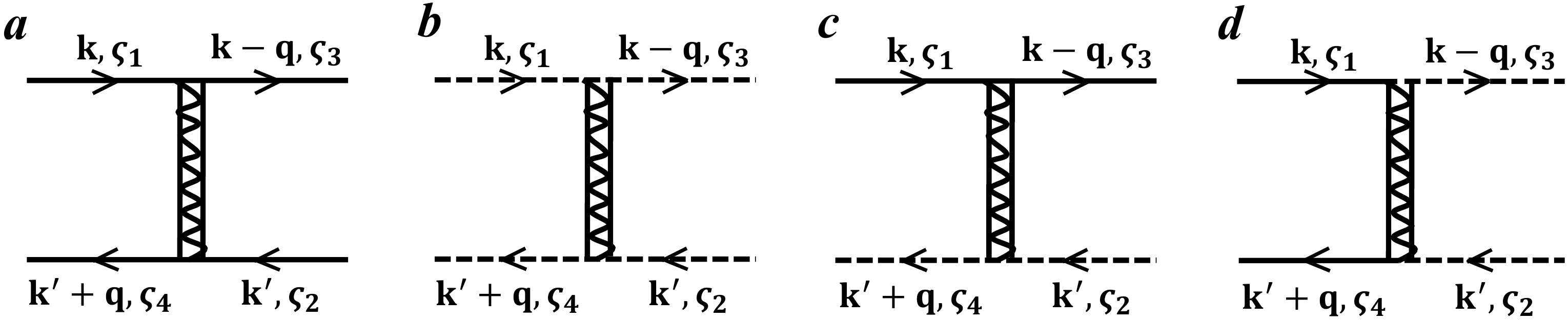}
\caption{\label{scatt} Interaction vertices for intra-valley ($a$ and $b$) and inter-valley ($c$ and $d$) scattering processes. The solid (dashed) lines refer to electrons in the $K$ ($K'$) valley. Diagrams $a$ and $b$ also have exchange partners with outgoing states swapped (not shown). Diagram $d$ involves a  large momentum transfer $\approx |\bf{K}-\bf{K}'|$. Reproduced with permission from Ref.~\onlinecite{Kumar_TMD:2021}. Copyright 2021 of the American Physical Society.}
\end{figure*}

 The valley degree of freedom in graphene allows for more interaction channels. In addition to direct and exchange interactions between electrons within a single valley, one now also has an exchange interaction between the valleys, and a mixed spin-valley exchange interaction. If doping is low enough to neglect the trigonal warping of the Fermi contours, the Landau function of a two-valley FL can be written as\cite{Raines_FL:2021}
 \bwt
 \bea
\label{FL}
N_F^*\hat  f(\bk,\bk')
&=&  
F^s(\vartheta) +
\bs
 \cdot\bs'
 F^a(\vartheta)
+
\btau_{||}
\cdot \btau_{||}'
G^{a||}(\vartheta)+\tau_{z}
\tau_{z}'
G^{a\perp}(\vartheta)
+ (\bs
\cdot \bs'
)\left[ \btau_{||}
\cdot \btau_{||}'
H^{||}(\vartheta)+\hat\tau_{z}
 \hat\tau_{z}'
 H^{\perp}(\vartheta)\right],
\eea
\ewt
where 
where $N_F^*$ is the total (including the spin and valley degeneracies), renormalized density of states, $\btau_{||}=\hat x\tau_x+\hat y\tau_y$ and unprimed/primed $\tau$-matrices refer to the valley state of the first/second quasiparticle. Scattering processes in a two-valley FL are depicted diagrammatically in Fig.~\ref{scatt}, where the solid (dashed) lines refer to electrons in the $K$ ($K'$) valley and  $\vs_{1,2}$ ($\vs_{3,4}$) label the spin indices of the initial (final) states.  Diagrams $a$ and $b$ describe intra-valley processes,  while diagram $c$ describes an inter-valley process with small momentum transfers, in which electrons stay in their respective valleys. Diagram $d$ corresponds to an inter-valley process with large ($\approx |\bf{K}-\bf{K}'|$) momentum transfer, in which electrons are swapped between the valleys. For Coulomb interaction, the corresponding matrix element is smaller than the matrix elements in diagrams $a-c$. If the inter-valley matrix element is neglected, then the rotational symmetry in the valley space is restored,\cite{Raines_FL:2021} i.e., $G^{a||}(\vartheta)=G^{a\perp}(\vartheta)$ and $H^{a||}(\vartheta)=H^{a\perp}(\vartheta)$. 

Collective modes of a FL in the presence of SOC and/or Zeeman magnetic field (i.e., a field that affects only electron spins but not their orbital motion) can be found from the self-consistent FL kinetic equation for the (matrix) occupation number $\hat n(\bk,\br,t)$
\bea
\label{kinetic_edsr}
\frac{\partial \hat{n}}{\partial t} + i [\hat{\ve}, \hat{n}]+\frac 12\left\{ \boldsymbol{\nabla}_\bk \hat \ve,  \cdot\boldsymbol{\nabla}_\br \hat n\right\} -\frac 12\left\{\boldsymbol{ \nabla}_\br \hat \ve, \cdot\boldsymbol{\nabla}_\bk \hat n\right\}  = 0,\nn\\
\eea
where $[\hat a_1,\hat a_2]=\hat a_1\hat a_2-\hat a_2\hat a_1$, $\left\{\hat{\bf a}_1,\hat{\bf a}_2\right\}=\hat{\bf a}_1\cdot\hat{\bf a}_2+\hat{\bf a}_2\cdot\hat{\bf a}_1$, and $\hat\ve(\bk,\br,t)$ is the functional of quasiparticle energy. In the most general case, $\hat\ve(\bk,\br,t)$ can be written as
\bea
\hat\ve(\bk,\br,t)=\hat\ve_0(\bk)+\hat\ve_{\text{SO}}(\bk)+\hat\ve_{\text{Z}}+\hat\ve_{\text{FL}}(\bk,\br,t),
\label{KE}
\eea
where $\hat{\ve}_0$ is the equilibrium quasiparticle energy in the absence of SOC and Zeeman field, $\hat\ve_{\text{SO/Z}}$ are the changes in the energy due to SOC/Zeeman field, and
\bea
\hat\ve_{\text{FL}}(\bk,\br,t)=
\text{Tr}' \int \frac {d^Dk'}{(2\pi)^D} \hat f(\bk,\bk') \hat n'(\bk',\br,t)
\eea
accounts for the interaction of a given quasiparticle with the rest. The effect of oscillatory magnetic and electric fields, applied in the ESR and EDSR measurements, are accounted for by adding the corresponding terms to the right-hand side (RHS) of \Eq{KE}. 

A number of comments are in order. 
\bi 
\item[i)]
In the case of graphene with broken A/B symmetry, the kinetic equation \eq{kinetic_edsr} needs to be modified to include the effect of a non-Abelian Berry curvature, which arises due to the combined effect of  broken A/B symmetry and Rashba SOC.\cite{raines:2022} In this case, the term $\boldsymbol{\nabla}_\bk\hat n$  is replaced by the covariant derivative $\mathcal{D} \hat n=\boldsymbol{\nabla}_\bk \hat n-i\left[\mathcal{A},\hat n\right]$, where $\mathcal{A}$ is the non-Abelian Berry connection.\cite{culcer:2005,xiao:2010,bettelheim:2017} This leads to an effective orbital magnetic field  which, in its turn, gives rise to an EDSR peak in the Hall conductivity, see Sec.~\ref{sec:GR}.
\item[ii)]Equation \ref{kinetic_edsr} neglects scattering of quasiparticles either by external sources (disorder, phonons) or by other quasiparticles, which lead to damping of collective modes. These effects will be discussed separately in Sec.~\ref{sec:damping}.
\item[iii)] The FL theory cannot describe a collective spin excitation that condenses out of the continuum of spin-flip transitions between the lower and upper cones of the Dirac surface state described by \Eq{HTI}. The reason is that the energy of such an excitation is comparable to $2E_F$, while the FL theory can only describe physics at energies much smaller than $E_F$. In this case, one has to use microscopic, rather than phenomenological methods to describe the electron-electron interaction, see Sec.~\ref{sec:Bi2Se3}.
\item[iv)] If the reader is content with our assumption that the effects of SOC and Zeeman field on a FL can be treated perturbatively, they can skip the next section and go directly to Sec.~\ref{sec:Silin}. A more demanding reader is invited to read the next section, which explains why the FL theory cannot deal with strong SOC and/or strong Zeeman field.
\ei
\subsection{Does the Fermi-liquid theory work for arbitrarily strong spin-orbit coupling and/or magnetic field?}
\label{sec:noFL}
\begin{figure}[htb]
\centering
\includegraphics[width=1\columnwidth]{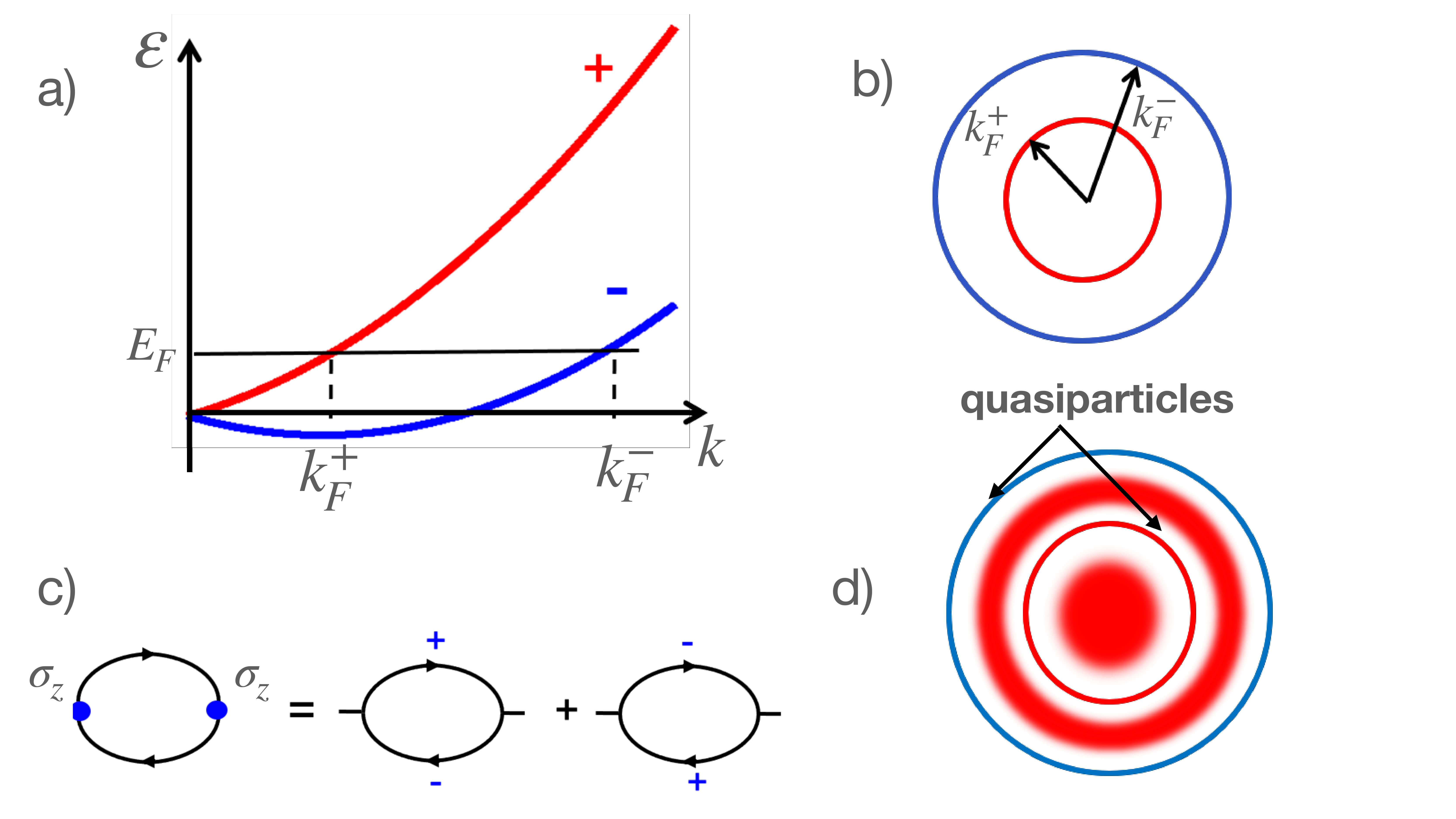}
\caption{a) Spectrum of the Rashba Hamiltonian \eq{HR}. b) The Fermi surface of the Rashba-split spectrum for non-interacting electrons. c) Diagrammatic representation of the out-of-plane spin susceptibility, $\chi_{zz}$. $\pm$ label the Green's functions of the Rashba subbands. d) Same as in b) for interacting electrons. Red shaded regions denote states away from the Fermi circles which cannot be described by the Fermi-liquid theory.}
\label{fig:Rashba_spectrum}
\end{figure}

At first glance, the answer to the question in the title of this section is in the affirmative. All one needs to do is to construct a new Landau interaction function, accounting for broken rotational invariance in the spin space and, in the case of SOC, for coupling between momentum and spin. A modification of the Landau function in \Eq{FSU2} for the case of a ferromagnetic metal was proposed long time ago in Ref.~\onlinecite{abrikosov:1958}. For the case of Rashba SOC, the modified Landau function was composed in Ref.~\onlinecite{ashrafi:2013}, but it is too long to be displayed here. It suffices to say that it contains eight instead of two components which, in contrast to  \Eq{FSU2}, depend not only on the angle between $\bk$ and $\bk'$ but also on the magnitudes of these momenta. And this is the first sign of a problem with the FL theory. To make this problem more evident, we consider the case of non-interacting electrons with Rashba spectrum (cf. Fig.~\ref{fig:Rashba_spectrum}a):
\bea
\ve^{\pm}_k=\frac{k^2}{2m}\pm \alpha k.
\eea
If both Rashba branches of the spectrum are occupied, the Fermi surface consists of two concentric circles with radii $k_{F}^\pm=\mp m\alpha+\sqrt{(m\alpha)^2+2mE_F}$, as shown in Fig.~\ref{fig:Rashba_spectrum}b. Let us  calculate the spin susceptibility. In the diagrammatic language, the $ij$ component of the spin susceptibility is given by a polarization bubble with the corresponding Pauli matrices at the vertices.  In the absence of SOC, the spin susceptibility comes from the states in the immediate vicinity of the Fermi surface. This is not so in the presence of SOC. The simplest case is when the Zeeman magnetic field is applied along the normal to the 2DEG plane, i.e., $i=j=z$, see Fig.~\ref{fig:Rashba_spectrum}c. In this case the $T=0$ static susceptibility arises solely from transitions between the two Rashba branches: 
\bea
\chi_{zz}&
=&\frac{g^2\mu_B^2}{2}\int_0^\infty \frac{dk k}{2\pi}\frac{n_{F}(\ve^{-}_k)-n_{F}(\ve^+_k)}{\ve^+_k-\ve^-_k}\nn\\
&&=\frac{g^2\mu_B^2}{2}\int_{k_{F}^+}^{k_{F}^-} \frac{dk}{4\pi\alpha}=\frac{g^2\mu_B^2}{2}\frac{m}{2\pi},\label{chizz}
\eea
where $n_{F}(\ve)$ is the Fermi function. We see that the integral over $k$ does not come from the vicinity of either Fermi circle, but rather from the entire interval between the Ferm circles of width $\Delta k_F=k_{F}^--k_{F}^+=2m\alpha$.\footnote{The in-plane components, $\chi_{xx}=\chi_{yy}$, consist of two parts: one is the Ferm-surface contribution, as in the absence of SOC, and another one is the contribution from the entire interval $\Delta k_F$.} It is not a problem for a non-interacting case, because the spectrum is known for arbitrary $k$.\footnote{Note that the integral in \Eq{chizz} can be also solved by rewriting it as $\chi_{zz}=(1/4\pi\alpha)\int^\infty_0 dk\left[n_{F}(\ve^{+}_k)-n_{F}(\ve^-_k)\right]$ and then integrating by parts. In this way, one obtains the same result as the sum of two contributions from the Fermi circles. This means that spin susceptibility is an anomalous quantity in the field-theoretical sense, i.e., it can be viewed equivalently either as low-energy or high-energy contribution.\cite{anomaly:book} But this equivalence is lost in the presence of interactions.} But it does become a problem for the interacting system. Indeed, the concept of quasiparticles is applicable only to long-lived states in the vicinity of each of the Fermi circle, see  Fig.~\ref{fig:Rashba_spectrum}d. But states away from each of the Fermi circles (shown by red shaded regions) are just some complicated many-body states, which the FL theory cannot describe.

Suppose that one ignores this warning and goes ahead with calculating the renormalized $g$-factor, using the Landau function modified by Rashba SOC from Ref.~\onlinecite{ashrafi:2013}. The only modification which matters here is that, because SOC breaks spin-rotational invariance, the $\bs\cdot\bs'F^a$ term in \Eq{FSU2} is replaced by a combination $\bs_{||}\cdot \bs'_{||}f^{a||}(\bk,\bk')+\hat\sigma_{z}\hat\sigma'_{z}f^{a\perp}(\bk,\bk')$, where $\bs_{||}=\hat\sigma_x\hat{\bf x}+\hat\sigma_y\hat{\bf y}$. With this modification, one follows the same procedure as for a usual FL \cite{agd:1963,lifshitz:1980} and arrives at the integral equation, for e.g., the out-of-plane $g$-factor
\bea
g_{zz}^*(k)=g-\int^{k^{*-}_{F}}_{k^{*+}_{F}} \frac{dk' k'}{\pi} \frac{f^{a\perp}_0(k,k')}{\ve^{*+}_{k'}-\ve^{*-}_{k'}}g_{zz}^*(k'),\nn\\\label{gfactor}
\eea 
where $^*$ indicates a renormalized quantity. (Note that the Luttinger theorem guarantees only that the total area of the Fermi surface is not renormalized, i.e., that $k^{+*2}_{F}+k^{-*2}_{F}=k^{+2}_{F}+k^{-2}_{F}$, which does exclude  renormalizations of $k_{F}^\pm$  individually.) 

A brief inspection of \Eq{gfactor} shows that, in general, it does not make sense. Indeed, the integral on its RHS involves energies of quasiparticles at an arbitrary point between two Fermi circles. However, as we explained before, quasiparticle states are not well-defined away from either of the Fermi circles. Therefore, the renormalization of the $g$-factor cannot be calculated within the FL theory. Only if SOC is weak and thus $\Delta k_F\ll k_F$, one can replace  $\ve^{*\pm}_k$ by their quasiparticle forms. But in this case the width of the integration interval cancels out with the energy splitting due to SOC, and we are back to the usual result for the renormalized $g$-factor: $g^*=g/(1+F^{a}_0)$. (For weak SOC, the difference between the in-plane and out-of-plane components of the exchange interaction becomes negligible, and we replaced $F^{a,\perp}_0$ by $F^{a}_0$ at the last step.)

A similar conclusion had eventually been reached in regard to a FL in the Zeeman magnetic field: if the Zeeman splitting is comparable to the Fermi energy, the FL theory does not work.\cite{meyerovich:1992a,meyerovich:1994,meyerovich:1994b,mineev:2004} Interestingly, problems with applying the FL theory to the cases of strong spin polarization, ferromagnetism, and SOC had been foreseen by Conway Herring as early as in 1966.\cite{herring:book} The problem discussed above is a generic feature of systems with broken $SU(2)$ symmetry. 

What was said above does not imply that an interacting system of electrons with strong SOC belong to the category of non-Fermi liquids. Indeed, momentum-resolved probes, such as angular-resolved photoemission, would find well-defined quasiparticles near each of the Fermi circles. What it means is that the thermodynamic quantities characterizing the spin sector of such a FL cannot be described by a (small) set of Landau parameters,  but must involve the information about states away from the Fermi surfaces. 

This is not such an unusual situation. For example, renormalization of the effective mass of a Galilean-invariant FL is described by a single Landau parameter: $m^*=m(1+F_1^s)$. However, if the system is not Galilean-invariant (but still isotropic), the last formula changes to $m^*=mQ(1+F_1^s)$, where $Q$ involves integrals of the interaction vertices over the entire momentum space, \cite{baym:book,chubukov:2018} and thus the effective mass in this case cannot be described by any finite number of Landau parameters. 

\section{Preamble: Silin modes in a partially spin-polarized Fermi liquid}
\label{sec:Silin}
\begin{figure}[H]
\centering
\includegraphics[width=0.8\columnwidth]{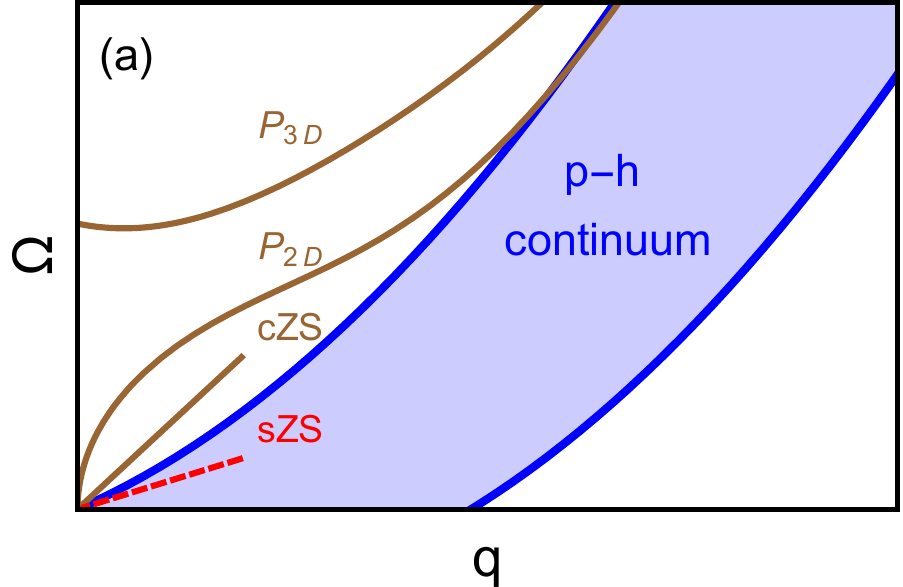}\\
\vspace{0.1in}
\includegraphics[width=0.8\columnwidth]{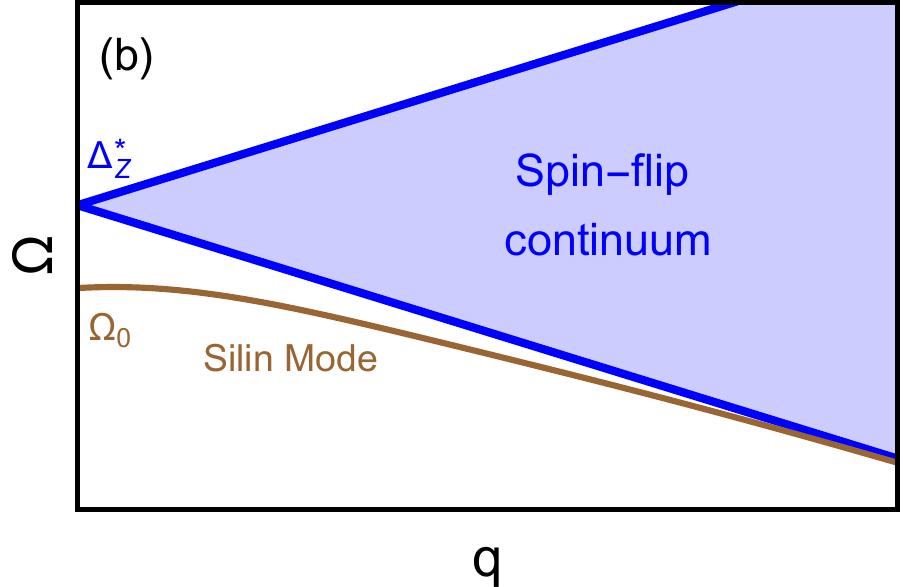}\\
\vspace{0.1in}
\includegraphics[width=0.8\columnwidth]{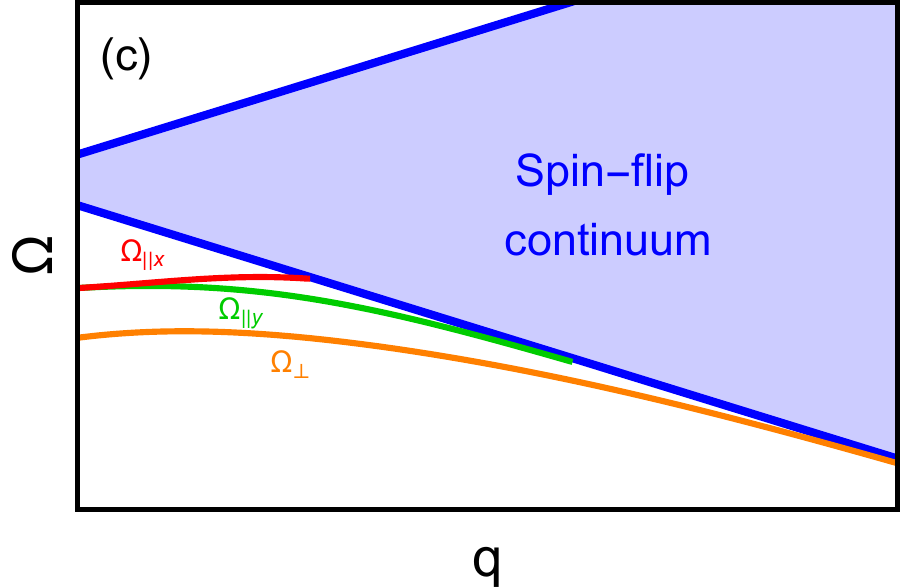}
\caption{\label{fig:collmodes_B0} a) Collective modes of an $SU(2)$-invariant Fermi liquid. cZS and sZS denote acoustic zero-sound modes of a neutral Fermi liquid in the charge and spin sectors, respectively. $P_{2D}$ and $P_{3D}$ denote plasmons of a charged Fermi liquid in 2D and 3D, respectively. b) Dispersion of the $n=0$ Silin mode in the presence of the magnetic field. c) Collective modes of a single-valley 2D Fermi liquid with Rashba and/or Dresselhaus spin-orbit coupling. $\Omega_{||,x}$ and $\Omega_{||,y}$ denote the frequencies of two modes with in-plane magnetizations, $\Omega_{\perp}$ is the frequency of the out-of-plane mode.}
\end{figure}

Switching to the subject of collective modes, it is instructive to begin with a short reminder about the collective modes in a single-valley FL without SOC.\cite{silin:1958,lifshitz:1980,baym:book} 
A neutral FL, e.g., normal $^3$He, supports long-lived collective excitation in the (charge) density sector: zero-sound waves with acoustic spectrum (line cZS in Fig.~\ref{fig:collmodes_B0}$a$). In a charged FL, the $l=0$ branch of zero-sound waves a replaced by plasmons, which are gapped in 3D (curve $P_{3D}$) and gapless in 2D (curve $P_{2D}$). The spin zero-sound mode does not fare that well. There is no general theorem that such a mode cannot be long-lived. However,  as long as the interaction between fermions is repulsive (which corresponds to the attractive exchange interaction) and under reasonable assumptions about the harmonics of the Landau function $F^a(\vartheta)$ in \Eq{FSU2}, i.e., $F_0^a<F_1^a<0$, etc., the mode is subsumed by the particle-hole continuum and thus overdamped (line sZS in Fig.~\ref{fig:collmodes_B0}$a$).

The situation changes in the presence of the Zeeman magnetic field, which gaps out the continuum of particle-hole excitations accompanied by spin-flips, see Fig.~\ref{fig:collmodes_B0}$b$. At $q=0$ the gap in the continuum is equal to the renormalized Zeeman energy $\dz^*=\dz/(1+F_0^a)$, then it decreases as $q$ increases, and vanishes eventually at $q\approx \dz/v_F^*$, where $v_F^*$ is the renormalized Fermi velocity. The region below the lower boundary of the continuum is now free of particle-hole excitations and can support long-lived collective modes. 

The frequencies of these modes at $q=0$  are found by writing the occupation number as $\hat n(\bk,t)=n_F+n_F'\left[\dz^*/2+{\bf u}(\bk,t)\cdot\bs\right]$
and the quasiparticle energy as $\hat\ve(\bk,t)=\ve_0+\dz^*\hat {\bf b}\cdot\bs/2+\hat\ve_{\text{FL}}(\bk,t)$, where $\hat {\bf b}$ is the unit vector in the direction of the magnetic field and $n_F'=\partial n_F(\ve)/\partial\ve$, and using 
 \Eq{kinetic_edsr} without the gradient term.\cite{lifshitz:1980,baym:book} This gives an equation of motion for the vector ${\bf u}$
 \bea
 i\partial_t {\bf u}(\bk_F,t)=\dz^* \hat {\bf b}\times\left[ {\bf u}(\bk_F,t)+\int\frac{d\mathcal{O}'}{\mathcal{O}_D} {\bf u}(\bk_F',t)F^a(\vartheta)\right],\nn\\
 \label{silin_eq}\eea
 where $\bk_F=k_F \bk/k$ and $\bk_F'=k_F \bk'/k'$, $d\mathcal{O}'$ is the element of the solid angle subtended by the vector $\bk_F'$, and 
$ \mathcal{O}_D$ is the full solid angle in $D$ dimensions. Equation \eqref{silin_eq}, which describes precession of the vector ${\bf u}$ around the magnetic field, is an eigenmode equation for the Silin collective modes.\cite{silin:1958} The frequencies of these modes depend on the angular momentum,  $\ell$, in 3D  or its projection onto the normal to the plane of motion, $m$,  in 2D \cite{baym:book}
\bea
\Omega_n=\dz^*(1+F^{a}_n)=\dz \frac{1+F^{a}_n}{1+F^{a}_0},\label{silin_freq}
\eea
with $n=\ell$ in 3D and $n=m$ in 2D. Here, $F^{a}_n$ are the harmonics of the Landau function defined by
\bea
F^a(\vartheta)=\sum_{\ell} (2\ell +1) F_\ell^a \mathcal{P}_\ell(\vartheta) 
\eea
and
\bea
F^a(\vartheta)=\sum_{m} F_m^a e^{im\vartheta}
\eea
in 3D and 2D, respectively, and $\mathcal{P}_\ell(\vartheta)$ are the Legendre polynomials. Although there is, in principle, an infinite number of Silin modes, only the $n=0$ mode couples to an oscillatory magnetic field, applied in an ESR or nuclear magnetic resonance (NMR) measurements.   Indeed, the magnetization is expressed solely through the zeroth harmonic of the function ${\bf u}(\bk_F,t)$: 
$\boldsymbol{\mathcal{S}}=-(1/2)g^*\mu_BN_F^* \int d\mathcal{O} {\bf u}(\bk_F,t)/\mathcal{O}_D=(1/2)g^*
\mu_BN_F^*{\bf u}^0(t)$. 

As we see from \Eq{silin_freq}, the frequency of the $n=0$ mode is not renormalized by the interaction and coincides with the Larmor frequency for free fermions. This is a general property of Hamiltonians with interactions that conserve spin and do not depend on velocities.\cite{ma:1968} Diagrammatically, the results comes about as a cancellation between the self-energy and vertex corrections the spin susceptibility.\cite{ma:1968}

\begin{figure}[htb]
\centering
\includegraphics[width=1.0\columnwidth]{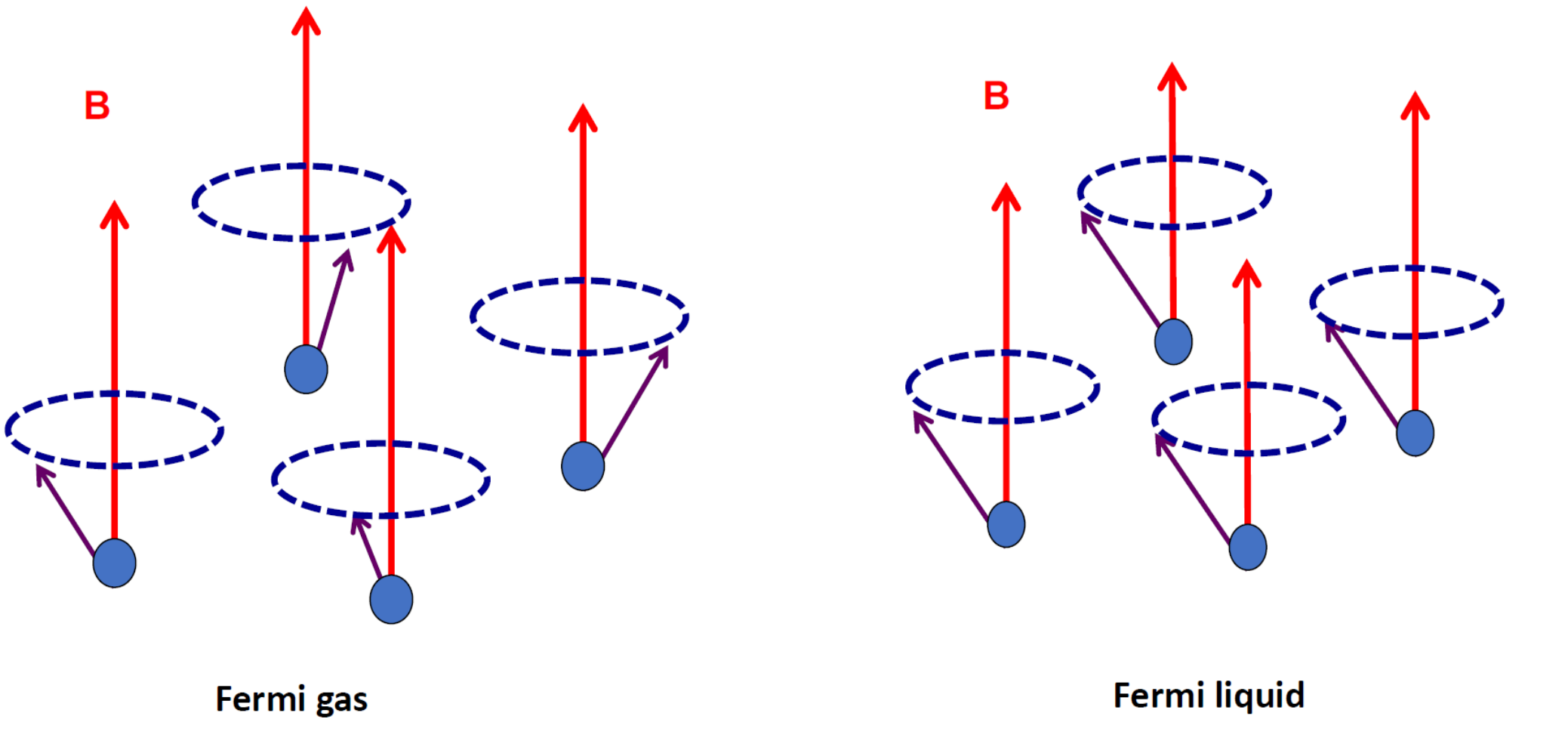}
\caption{Precession of spins in a Fermi gas (left) and Fermi liquid (right). Although the frequencies of the $n=0$ mode are the same in both cases, the phases of precessing spins are uncorrelated in a Fermi gas and locked in a Fermi liquid.}
\label{fig:precession}
\end{figure}

The difference between spin precession of free fermions and FL quasiparticles is illustrated in Fig.~\ref{fig:precession}. Although the spins of free fermions (on the left) and FL quasiparticles (on the right) precess with the same frequencies, the phases of the former are not correlated but the phases of the latter are locked.

The $n=0$ Silin mode disperses down with $q$, at first quadratically, and then grazes the continuum of spin-flip excitations, as shown in Fig.~\ref{fig:collmodes_B0}$b$. The downward sign of the dispersion can be explained by attraction in the exchange channel of the interaction.\cite{perez:2007} Unlike the frequency at $q=0$, which is not renormalized by the interaction, the functional form of the dispersion encapsulates all harmonics of $F^a$. In particular, the quadratic part contains harmonics $F^a_0$ (which can be also extracted from the spin susceptibility, if the effective mass is measured independently from the specific heat) and $F_1^a$ (which cannot). The dispersion of the Silin mode in normal $^3$He was studied extensively by NMR experiments in a spatially varying magnetic field which produced confined waves, see Ref.~\onlinecite{candela:1986} and references therein. In solid-state systems, the Silin modes where measured in the late 60s by ESR on alkali metals. \cite{schultz:1967,platzman:1967} More recently, the dispersion of the Silin mode was measured by high-precision, finite-$q$ Raman spectroscopy in a Cd$_{1-x}$Mn$_{x}$Te quantum wells.\cite{perez:2007, baboux:2013,baboux:2015,perez:2016,perez:2017,perez:2019} In this case, however, the Silin mode bears clear fingerprints of both Rashba and Dresselhaus SOCs, which are discussed  in Sec.~\ref{sec:finiteq}. A tantalizing proposal to observe Silin modes in a quantum spin liquid with spinon Fermi surface has recently been put forward in Ref.~\onlinecite{starykh:2020}.

\section{Chiral spin waves in two-dimensional electron gases}
\label{sec:2DEG}
\subsection{Chiral spin waves with Rashba spin-orbit coupling}
\label{sec:RSOC}
We now proceed with discussing the main subject of this paper--collective spin modes in the presence of SOC--starting from the simplest case of a single-valley 2DEG with Rashba SOC, studied first in Ref.~\onlinecite{shekhter:2005}. The single-particle Hamiltonian is given by \Eq{HR} and the Landau function is given by \Eq{FSU2}. It is convenient to introduce a set of rotated Pauli matrices which depend on the electron momentum\cite{shekhter:2005}
\bea
\hat\Sigma_{x}(\bk)&=&-\hat\sigma_z,\;\hat\Sigma_y(\bk)=\cos\phi_\bk\hat\sigma_x+\sin\phi_\bk\hat\sigma_y,\nn\\
\hat\Sigma_z(\bk)&=&\sin\phi_\bk\hat\sigma_x-\cos\phi_\bk\hat\sigma_y,\label{Sigma}
\eea
where $\phi_\bk$ is the azimuthal angle of $\bk$. When taken at the same momentum, $\hat\Sigma$'s obey the usual algebra $\left[\hat\Sigma_{\alpha}(\bk),\hat\Sigma_\beta(\bk)\right]=i\epsilon^{\alpha\beta\gamma}\hat\Sigma_\gamma(\bk)$, where $\epsilon^{\alpha\beta\gamma}$ is the Levi-Civita tensor, while $\left[\hat\Sigma_{y}(\bk),\hat\Sigma_z(\bk')\right]=i\hat\Sigma_x(\bk)\cos(\phi_\bk-\phi_{\bk'})$. In this basis, 
\bea
\hat\ve_{\text{SO}}=\frac 12 \dr \hat\Sigma_{z}(\bk),
\eea
where $\dr=2\alpha k_F$. Treating $\hat\ve_{\text{SO}}$ as a static perturbation, we obtain the renormalized value of the Rashba splitting\cite{raikh:1999,shekhter:2005}
\bea
\dr^*=\dr/(1+F^a_1).\label{renorm}
\eea 
This energy marks the end point of the continuum of particle-hole excitations which involve spin-flip transitions between the branches of the Rashba spectrum, see Fig.~\ref{fig:collmodes_B0}$c$. 
In a non-interacting system, the continuum at $q=0$ occupies a finite region 
of width $\Delta\Omega=4m\alpha^2$. However, the FL theory is valid only to first order in $\alpha$, therefore, within this theory the continuum shrinks to a single point at $\Omega=\dr^*$.

Now we consider a time-dependent perturbation of the occupation number
\bea
\delta\hat n(\bk,t)=n_F'\left[\frac 12 \dr^* \hat\Sigma_{z}(\bk)+\bu(\bk,t)\cdot\hat{\boldsymbol{\Sigma}}(\bk)\right].\nn\\
\eea 
Substituting this form into \Eq{kinetic_edsr} (without the gradient term) and linearizing with respect to $\bu$, we obtain the following equations of motion for angular harmonics of $\bu$:
\begin{subequations}
\bea
\partial_t u_{x}^m&= &-\dr^* u_{y}^m \left[1 +\frac{1}{2} \left(F_{m-1}^a + F_{m+1}^a \right) \right],
\label{u1_m}\\
\partial_t u_{y}^m&=& \dr^*u_{y}^m \left( 1+F_{m}^a \right),
\label{u2_m}\\
\partial_t u_{3}^m&=&0.
\label{u3_m}
\eea
\end{subequations}
The last equation implies that $u_{3}^m=\text{const}$ and can thus be ignored.
The eigenvalues of this system are the frequencies of a new type of collective modes: chiral spin resonances.\cite{shekhter:2005} Their frequencies are
\bea
 \Omega_m=\dr^*\sqrt{\left[1+\frac 12(F_{m+1}^a + F_{m-1}^a)\right](1+F_m^a)}.\nn\\
 \label{freqR}
 \eea
 (Note that $\Omega_{-m}=\Omega_m$ because $F^a_{-m}=F^a_m$.)
 As in the case of the Silin mode, there is an infinite number of chiral-spin resonances. Now, however, only two modes with $m=0$ and $m=1$ couple to the external magnetic field. This is because the magnetization 
 \bea
 \boldsymbol{\mathcal{S}}=-\frac{1}{4} g^*\mu_B N_F^*\text{Tr}\int \frac{d\phi_\bk}{2\pi} \left[\bu(\bk_F,t)\cdot\hat{\boldsymbol{\Sigma}}(\bk_F)\right]\bs
 \eea
contains not only the $0^{\text{th}}$ but also $\pm 1^{\text{st}}$ harmonics of $\bu$ due to the angular dependence of matrices $\hat\Sigma_\alpha(\bk)$. The $m=0$ mode with frequency
\bea
\Omega_0\equiv\Omega_\perp=\dr^*\sqrt{(1+F_0^a)(1+F_1^a)}\label{Operp}
\eea
corresponds to oscillations of the $z$-component of $\boldsymbol{\mathcal{S}}$, while the doubly-degenerate  $m=\pm 1$ mode with frequency
\bea
\Omega_{\pm 1}\equiv\Omega_{||}=\dr^*\sqrt{\left[1+\frac 12\left(F_0^a+F_2^a\right)\right]\left(1+F_1^a\right)}\label{Opar}
\eea
corresponds to oscillations of the in-plane components of $\boldsymbol{\mathcal{S}}$. Provided that $F_0^a<F_2^a$, we have $\Omega_\perp<\Omega_{||}$, which is the case shown in Fig.~\ref{fig:collmodes_B0}$c$. Since spin is not a conserved quantity anymore, the frequencies of all modes are renormalized away from $\dr$. 

The chiral spin resonances are the $q=0$ end points of the dispersive modes: chiral spin waves (CSW).\cite{ashrafi:2012,maiti:2014}  Once the direction of $\bq$ in the plane is chosen, the degeneracy of the two  in-plane mode is lifted and there are now two modes with frequencies $\Omega_{||x}(\bq)$ and $\Omega_{||y}(\bq)$, polarized along and perpendicular to $\bq$, respectively. The dispersions of these modes are shown schematically in Fig.~\ref{fig:collmodes_B0}$c$. The in-plane modes run into the continuum at some values of $q$ and disappear from the spectrum. The out-of-plane mode behaves similarly to the Silin mode: it disperses down with $q$ and grazes the continuum of particle-hole excitations. We will discuss the spatial dispersion of CSW in more detail in Sec.~\ref{sec:space}.

\begin{figure}[htbp]
\centering
\includegraphics[scale=0.26]{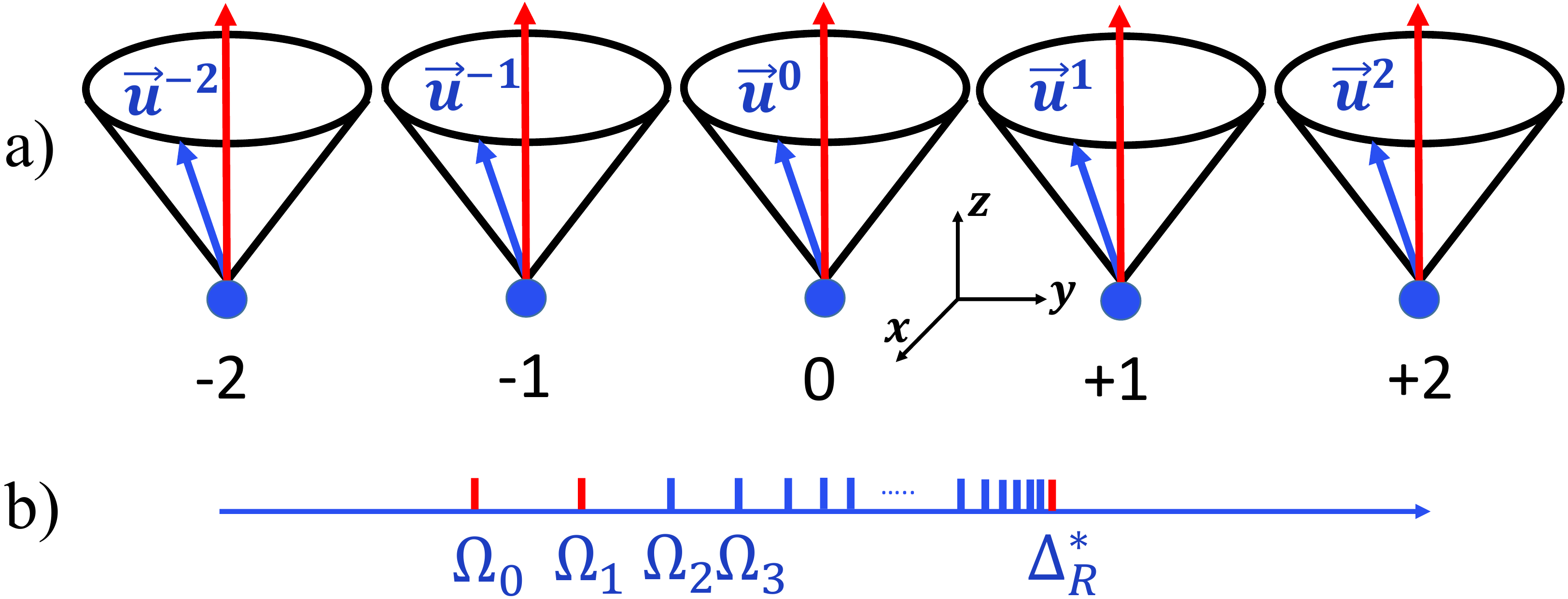}
\caption{\label{ESR_tightbinding} a) An interpretation of collective modes in a Fermi liquid with Rashba spin-orbit coupling in terms of a linear spin chain. b) The spectrum of the system consists of an infinite number of discrete levels, converging towards the continuum at $\dr^*$. Reprinted with permission from Ref.~\onlinecite{kumar_lattice:2017}. Copyright 2017 by the American Physical Society.}
\end{figure}

The equations of motion \eq{u1_m}-\eq{u3_m} allow for a simple physical interpretation.\cite{kumar_lattice:2017}
Namely, one can think of vector ${\bf u}^m(t)$ as a classical spin on site $m$ of a linear spin chain, aligned along the $y$-axis as in Fig.~\ref{ESR_tightbinding}.  Spins do not interact with each other but are subject to an effective magnetic field due to Rashba SOC, directed along the $z$ axis and of magnitude $\dr^*/2$.  The effective Land{\'e} factor of these spins is anisotropic in the $(x,y)$ plane with components $\gamma^m_x=2+ F_{m+1}^a+F_{m-1}^a$ and $\gamma^m_y=2(1+F_m^a)$.
The spin chain is {\em non-uniform} because $\gamma_1^m$ and $\gamma_2^m$ depend on the lattice site.  Both anisotropy and site-dependence  of the $g$-factor arise from the FL interaction. Because $u^m_3=\text{const}$,
the spins precess around the Rashba field.
The effective-lattice interpretation will be even more instructive in the case of both Rashba SOC and Zeeman field being present, which is discussed in the next section.

\subsection{Effective lattice model\label{sec:lattice}}

Now we add a Zeeman field aligned along the $x$-axis. The total Hamiltonian is the sum of Eqs.~\eq{HR} and \eq{HZ}, and the additional term in the quasiparticle energy in \Eq{KE} is
\bea
\hat\ve_{\text{Z}}=\frac{\dz^*}{2} \hat\sigma_x=\frac{\dz^*}{2}\left[\cos\phi_\bk\hat\Sigma_x(\bk)+\sin\phi_\bk\hat\Sigma_y(\bk)\right].\nn\\
\label{SZ}
\eea
Correspondingly, the equations of motion for vector $\bu$ become
\bwt
\begin{subequations}
\bea
\partial_t u_{x}^m &= &-\dr^* u_{y}^m \left[ 1 +\frac{1}{2} \left(F_{m-1}^a + F_{m+1}^a \right) \right] - \frac{1}{2i}\dz^*\left(u_{y}^{m-1} - u_{y}^{m+1} \right) \left( 1+F_{m}^a \right)
\nn\\
&&+ \frac{1}{2i}\dr^*u_{z}^m \left( F_{m-1}^a - F_{m+1}^a \right) + \frac{1}{2}\dz^*\left( u_{z}^{m-1} + u_{z}^{m+1} \right)\left( 1+F_{m}^a \right),
\label{u1_mZ}\\
\partial_t u_{y}^m &=& \dr^*u_{x}^m \left(1+F_{m}^a \right) + \frac{1}{2i}\dz^* \left[ u_{x}^{m-1} \left(1+F_{m-1}^a \right) - u_{x}^{m+1} \left(1+F_{m+1}^a \right) \right],
\label{u2_mZ}\\
\partial_t u_{3}^m &=& -\frac{1}{2} \dz^* \left[ u_{x}^{m-1} \left(1+F_{m-1}^a \right) +  u_{1}^{m+1} \left(1+F_{m+1}^a \right) \right].
\label{u3_mZ}
\eea
\end{subequations}
\ewt
Note that the angular dependence of the Zeeman energy in the $\Sigma$-basis in \Eq{SZ} leads to a non-locality in the $m$-space: the Zeeman terms in the equation of motion shift the angular momentum by $\pm 1$.  Now the equations of motion resemble those for the tight-binding (TB) model  with three orbitals per site, in which the Rashba and  Zeeman terms play the roles of on-site and hopping energies, respectively. To make this analogy more transparent, we eliminate components $u_y^m$ and $u_z^m$ and introduce the ``Bloch wavefunction'' 
\bea
\psi_m=i^{-m}u^m_x.\label{psi}
\eea
 Assuming the oscillatory time dependence with frequency $\Omega$, the equation for $\psi_m$ is reduced to the TB form:\cite{kumar_lattice:2017}
\begin{widetext}
\bea
\label{TB}
\Omega^2\psi_m&=& \left[\dr^{*2} \left( 1+F_m^a \right) \left( 1+\frac{F_{m+1}^a + F_{m-1}^a}{2} \right) + \dz^{*2} \left( 1+F_m^a \right)^2 \right] \psi_m\nn \\
&&-
\dr^{*} \dz^{*}
\left[
\left( 1 + \frac{F_{m}^a + F_{m+1}^a}{2} \right) \left( 1 + F_{m+1}^a \right) 
\psi_{m+1} +
\left( 1 + \frac{F_{m-1}^a + F_m^a}{2} \right) \left(1 + F_{m-1}^a \right) 
 \psi_{m-1}\right].
\eea
\end{widetext} 
In the absence of interaction,  \Eq{TB} is simplified to
\beq
\Omega^2\psi_m=\left(\dr^2+\dz^2\right)\psi_m-\dr\dz\left(\psi_{m+1}+\psi_{m-1}\right).\label{v1mB}
\eeq
The eigenvalue of this equation 
\bea
\Omega(\phi_\bk)=\left[\dr^2+\dz^2-2\dr\dz
\cos\phi_\bk\right]^{1/2}
\label{free}
\eea
is nothing but the difference between the energies of the Rashba subbands at a given direction of $\bk$ (modulo a phase shift inflicted by the transformation \eq{psi}).
$\Omega(\phi_\bk)$ disperses with $\phi_\bk\in(0,2\pi)$, which is a conjugate variable to $m$. Therefore, $\phi_\bk$ plays the role of ``quasimomentum" confined to the first Brillouin zone $(0,2\pi)$. 
The minimum and maximum values of $\Omega(\phi_\bk)$, $\Omega_{\max}=\dr+\dz$ and $\Omega_{\min}=|\dr-\dz|$,
mark the edges of the ``conduction band'', which is nothing but the particle-hole continuum. At $\dz=\dr$ the gap in the continuum collapses to zero. At this value of the magnetic field, the Fermi contours of the Rashba branches touch at one point and thus a transition between the two branches costs no energy. 

We now come back to the interacting version of \Eq{TB}, which allows for a simple physical interpretation. Imagine a 1D lattice with sites labeled by index $m=0,\pm 1,\dots$.  The Bloch wavefunction $\psi_m$ resides on these sites.  The first term on the RHS of \Eq{TB} is  the energy on site $m$, the last two terms describe hopping between site $m$ and its the nearest neighbors, $m\pm 1$. If we artificially remove all $F_m^a$'s from the equation but keep the renormalized values of $\dr$ and $\dz$, we will obtain the renormalized band (continuum) with boundaries $\Omega^*_{\max}=\dr^*+\dz^*$ and $\Omega^*_{\min}=|\dr^*-\dz^*|$. The role of $F_m^a$'s is to renormalize both on-site energies and hopping matrix elements. Because $F_m^a$ depends on $m$, both the on-site energies and hopping matrix elements vary along the lattice. In other words, $F_m^a$'s introduce both one-site and bond defects.  

In principle, $F^a_m$ is non-zero for any $m$. Therefore, each site of our lattice is different from the others, and we can view our lattice as an ordered alloy composed from an infinite variety of chemically distinct atoms. But if we keep only a few first harmonics of $F^a(\vartheta)$ and neglect the rest, the central region of the lattice will contain defects, while the outer regions will be defect-free. Lattice defects produce bound states which split off the band. In this language, therefore, the collective modes of a FL  are the bound states of an effective 1D lattice.
Studying a much more transparent problem of bound states, one can understand a more complicated case of a FL with SOC and in the presence of the magnetic field.  

The case of Dresselhaus SOC can be analyzed in a similar way. According to \Eq{HD}, the corresponding term in the quasiparticle energy reads
\bea
\hat\ve_{\text{SO}}&=&\frac{\dd^*}{2}\left(\cos\phi_\bk\hat\sigma_x-\sin\phi_\bk\hat\sigma_y\right)\nn\\
&=&\dd^* \left[\cos 2\phi_\bk\hat\Sigma_y(\bk)+\sin 2\phi_\bk\hat\Sigma_z(\bk)\right],\nn\\
\eea
where $\dd^*=\dd/(1+F_1^a)$.
A double angle in this equation implies that the corresponding terms in the equations of motion shift the angular momentum by $\pm 2$. In the lattice interpretation, this corresponds to hopping between next-to-nearest neighbors. 
A complete dictionary of mapping between the FL kinetic equation and TB model  is given in Table~\ref{table:I}.
\begin{table*}
\caption{\label{table:I}Mapping of the Fermi-liquid kinetic equation onto an effective 1D tight-binding model.}
\centering
\begin{tabular}{|c|c|}
\hline
{\bf Fermi-liquid kinetic equation} & {\bf 1D tight-binding model} \\  
\hline
angular momentum $m$ & lattice site $m$\\
\hline
azimuthal angle of momentum $\bk$ ($\phi_\bk$) & quasimomentum\\
\hline
 $m^\text{th}$ harmonic of the spin part of the occupation number $u^m_{x}$ & Bloch wavefunction on site $m$\\
\hline
Rashba spin-orbit coupling & on-site energy\\
\hline
Zeeman splitting & nearest-neighbor hopping\\
\hline
Dresselhaus spin-orbit coupling & next-to-nearest neighbor hopping\\
\hline
harmonics of the Landau function & on-site and bond defects\\
\hline
continuum of spin-flip particle-hole excitations & conduction band\\
\hline
collective modes & bound states\\
\hline 
 \end{tabular}
\end{table*}

We note that the zero-sound modes of a 2D FL in the absence of SOC and Zeeman field can also be understood as bound states of a 1D lattice. In this case, hopping between sites arises due the gradient term in the kinetic equation, which also shifts the harmonic index by $\pm 1$. For example, the shear zero-sound waves of a 2D FL were analyzed in this way in Ref.~\onlinecite{sodemann:2019}.

\subsection{Effective lattice model for Rashba spin-orbit coupling and Zeeman magnetic field}
\label{sec:RSOCZ}
\begin{figure}
\centering
\includegraphics[scale=0.35]{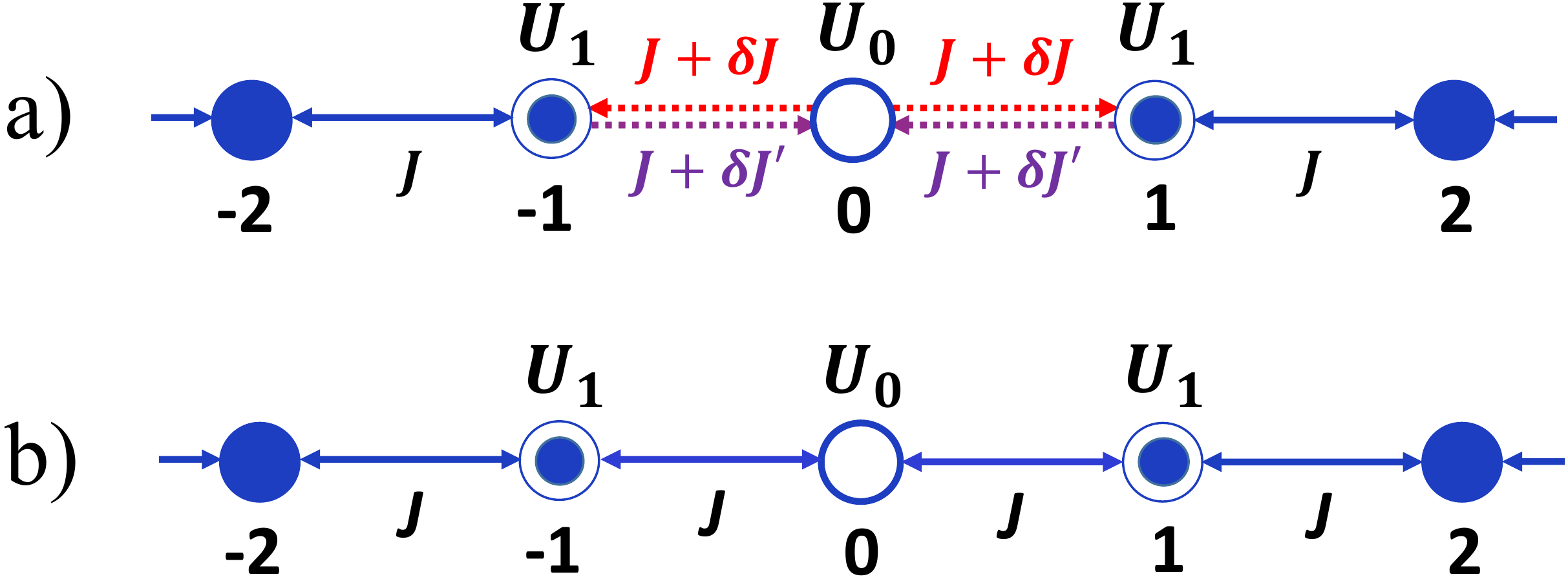}
\caption{\label{ref:chiral} a) Effective lattice for the FL kinetic equation \eq{tight binding s-wave} with Rashba spin-orbit coupling and in-plane magnetic field, and in the $s$-wave approximation for the Landau function.  Filled circles represent sites of ideal lattice with on-site energies $W$, connected by bonds with hopping amplitudes $J$. The $m=0,\pm 1$ sites are occupied by ``impurities'' with on-site energies $U_0$ and $U_{\pm 1}$, respectively, connected by ``defective'' bonds with hopping amplitudes $J\pm \delta J$. b) A simplified version of the effective lattice with on-site disorder only. Reprinted with permission from Ref.~\onlinecite{kumar_lattice:2017}. Copyright 2017 by the American Physical Society.}
\end{figure}

We now illustrate  how  the lattice model works  for the case when both Rashba SOC and Zeeman magnetic field are present. In addition, we assume that the Landau function is isotropic and thus contains only the zeroth harmonic: $F^a_m=\delta_{m,0}F_0^a$ (the $s$-wave approximation). In a dimensionless form  and in the $s$-wave approximation, Eq.~(\ref{TB}) can be written as
 \begin{widetext}
 \bea
 \omega^2 \psi_m&=&
 W\psi_m -J\left(\psi_{m+1}+\psi_{m-1}\right)+U_0\delta_{m,0}
 \psi_m
 +U_1\left(\delta_{m,1}+\delta_{m,-1}\right)
 \psi_m
- \delta J\delta_{m,0}
\left(\psi_1+\psi_{-1}\right)-
\delta J'\left(\delta_{m,1}+\delta_{m,-1}\right)
\psi_0,\nn\\
 \label{tight binding s-wave}
\eea
\end{widetext}
where 
\bse
\bea
\omega&=&\Omega/\dr,\;J=\dz^*/\dr,\;W=1+J^2,\label{par1}\\
U_0&=&F_0^a\left[1+J^2\left(2+F_0^a\right)\right],\;U_1=  \frac{F_0^a}{2}\label{par2}\\
\delta J&=&J \frac{F^a_0}{2},\;\delta J'=\frac{F^a_0}{2}\left(3+F_0^a\right).\label{par3}
\eea
\ese
 [Note that, according to Eq.~(\ref{renorm}),  $\dr$ is not renormalized in the $s$-wave approximation.]    The corresponding lattice is shown pictorially in Fig.~~\ref{ref:chiral}$a$. 
 The first two terms on the RHS of \Eq{tight binding s-wave} describe an ideal lattice with ``on-site energies'' $W$ and ``hopping amplitudes'' between the nearest neighbors $J$. The rest of the terms represent ``defects''. The third and fourth term correspond to three ``impurities'' on sites $m=0$ and $m=\pm 1$ with on-site energies $U_0$ and $U_1$, respectively. The last two terms describe defective bonds between the $0^{\text{th}}$ and $\pm 1^{\text{st}}$ sites.   The amplitudes are equal to $J+\delta J$ for hopping from $0$ to $\pm 1$ and  $J+\delta J'$ for hopping in the opposite direction, i.e., the bond defects are {\em chiral}. This means that the effective TB Hamiltonian is non-Hermitian. This does not present any difficulties, however,  because the eigenvalues of Eq.~(\ref{tight binding s-wave}) are real. 
  
\begin{figure*}[htbp]
\centering
\includegraphics[scale=0.5]{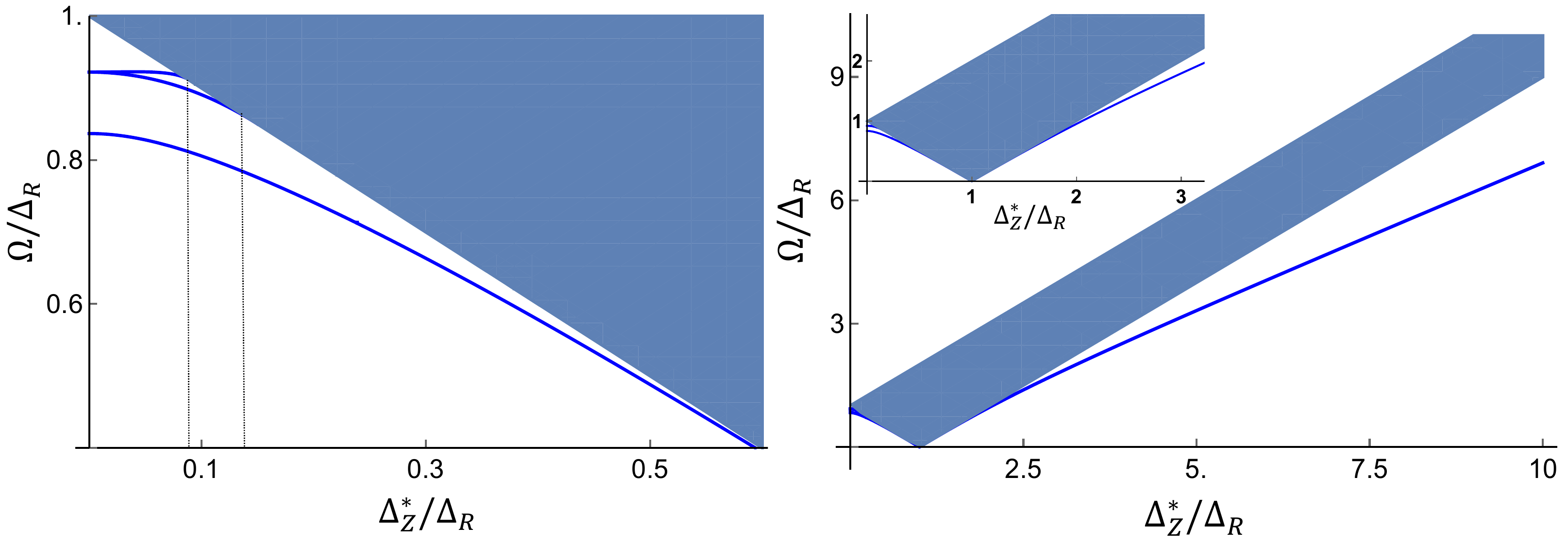}
\caption{\label{fig:R+B_s_wave} Collective modes of a FL with Rashba SOC and in the presence of the in-plane  magnetic field. The Landau function is taken in the $s$-wave approximation: $F^a(\theta)=F_0^a=-0.3$. $\Delta_Z^*$ is the (renormalized) Zeeman energy and $\dr$ is the Rashba energy splitting. Left: $\Delta_Z^*<\dr$. Right: $\Delta_Z^*>\dr$. Inset: Same as in main panel for a wider range of fields. Reprinted with permission from Ref.~\onlinecite{kumar_lattice:2017}. Copyright 2017 by the American Physical Society.}
\end{figure*}

 To solve  Eq.~(\ref{tight binding s-wave}),  we choose wavefunctions $\psi_0$ and $\psi_{\pm 1}$ as independent variables, and assume that, starting from sites $m=\pm 2$, the wavefunctions of the bound states decreases exponentially with $m$
 :\beq{\psi}_{\pm (|m|+2)} = e^{-(|m|+1)\lambda}
{\psi}_{\pm 1}
\label{Ansatz}
\eeq
 with $\text{Re}\lambda>0$. This yields a transcendental equation for $\lambda$, whose solutions are presented in Fig.~\ref{fig:R+B_s_wave}. The left panel is for magnetic fields below the gap closing point ($\dz^*<\dr$). There are two collective modes in zero field--these are the same modes as given by Eqs.~\eq{Operp} and \eq{Opar}. An in-plane magnetic field lifts the double degeneracy of the $\Omega_{||}$ mode (similar to the case of finite $q$ in Fig.~\ref{fig:collmodes_B0}$c$) and, at finite but small $\dz^*$ there are three modes.  The in-plane modes run into the continuum at some critical values of $\dz^*$,  but the out-of-plane mode continues to graze the continuum down the gap-closing point. For $\dz^*>\dr$, the single mode appears again, see Fig.~\ref{fig:R+B_s_wave}, right. Its frequency increases with the magnetic field and, in the limit of $\dz^*\gg \dr$, the mode evolves into the Silin mode with frequency $\dz$.
 
 A rather complex spectrum shown  in Fig.~\ref{fig:R+B_s_wave} can be understood in terms of simplified versions of the TB model. Namely, the merging of the two in-plane modes with the continuum can be understood qualitatively by ignoring bond defects, i.e., by setting $\delta J=\delta J'=0$. In this case, we have a TB model with identical bonds between all sites and three impurities on sites $m=0,\pm1$, see Fig.~\ref{ref:chiral}$b$. 
For an even simpler case of a single impurity in a 1D lattice, it is well-known that there always exists a bound state located either below (for an attractive impurity)  or above (for a repulsive impurity) the conduction band.  For realistic values $-1<F_0^a<0$, our ``impurities'' are attractive, and thus the bound state is below the band. Given that 
a single impurity has at least one bound state, it is natural to expect that a complex of three impurities will have up to three bound states, which is indeed confirmed by an explicit solution of the TB model (see Appendix A.1b in Ref.~\onlinecite{kumar_lattice:2017}). Qualitatively, this can be understood in the continuum limit, in which a three-impurity complex is replaced by a 1D potential well of finite width ($a$) and depth ($U$).  Such a well has at least one bound state  but may also have two, three, etc. states, 
if the product $Ua$ exceeds some critical values.   

Our original problem corresponds to a TB model with parameters given by  Eqs.~(\ref{par1}-\ref{par3}).
In the limit $\dz^*\ll \dr$ and not too weak interaction $|F_0^a|\sim 1$, the potential energies of the impurity sites are of the order of $1$, which is much larger than the bandwidth $2J=2\dz^*/\dr\ll 1$. Thus we have three strong impurities with the maximum number of bound states, equal to three. As $\dz^*$ increases, the bandwidth increases linearly whereas the potential energies increase only as $\dz^{*2}$. Therefore, the impurities get relatively weaker (compared to the bandwidth), and we lose first the highest and then next-to-highest bound state.  The lowest bound state also disappears but only at the gap-closing point. This seems to contradict the fact that there is at least one bound state in a 1D problem regardless to its parameters. However, this statement is true only for the continuum Schroedinger equation, which does not have the notion of bonds. In our full problem, two bonds are defective and there is an interesting competition between the on-site and bond defects, which does allow the bound state to disappear precisely at $\dz^*=\dr$.\cite{kumar_lattice:2017}

As the field increases beyond the gap-closing point ($\dz^*>\dr$), the $m=0$ impurity becomes stronger while the $m=\pm 1$ impurities remain the same. Therefore, we are back to a single-impurity problem with only one bound state or, equivalently, one collective mode.

\begin{figure}[htbp]
\centering
\includegraphics[width=\columnwidth]{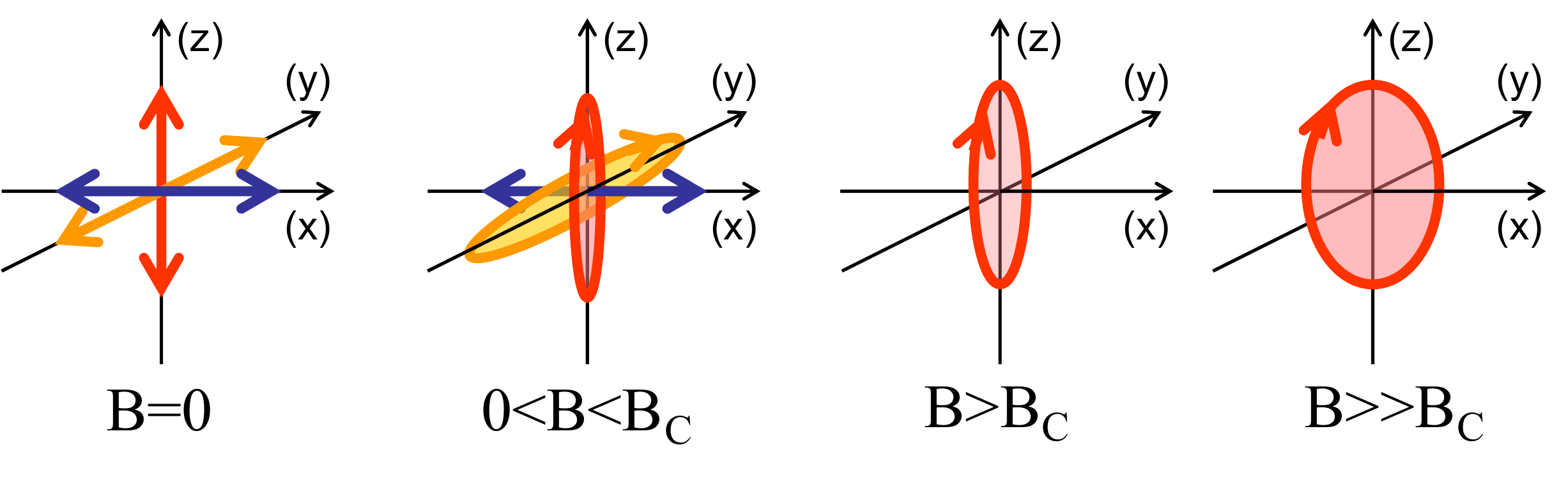}
\caption{\label{fig:polar} Polarizations of chiral spin modes at different values of the magnetic fields applied along the $x$-axis. $B_c$ is the gap-closing field, at which $\dz^*=\dr^*$. 
Reprinted with permission from Ref.~\onlinecite{maiti:2016}. Copyright 2016 by the American Physical Society.}
\end{figure}

From the solution of the kinetic equation one can also deduce the polarization of the collective modes. Figure \ref{fig:polar} shows how the polarization of chiral spin modes changes with the magnetic field.\cite{maiti:2016} In zero field, there are linearly polarized modes. At finite $B<B_c$, where $B_c$ is the field at which the gap in the continuum closes, along the $x$-axis, the longitudinal mode (with the magnetization along the field) remains linearly polarized, while two transverse mode (with the magnetizations perpendicular to the field) are elliptically polarized (as long as they are located outside the continuum). When the single wave emerges on the other side of the gap-closing point ($B>B_c$), it is elliptically polarized, with the magnetization vector precessing around the magnetic field. As the field becomes much larger than $B_c$, the wave transforms gradually into the Silin mode with circular polarization.

\begin{figure*}
\centering
\includegraphics[scale=0.5]
{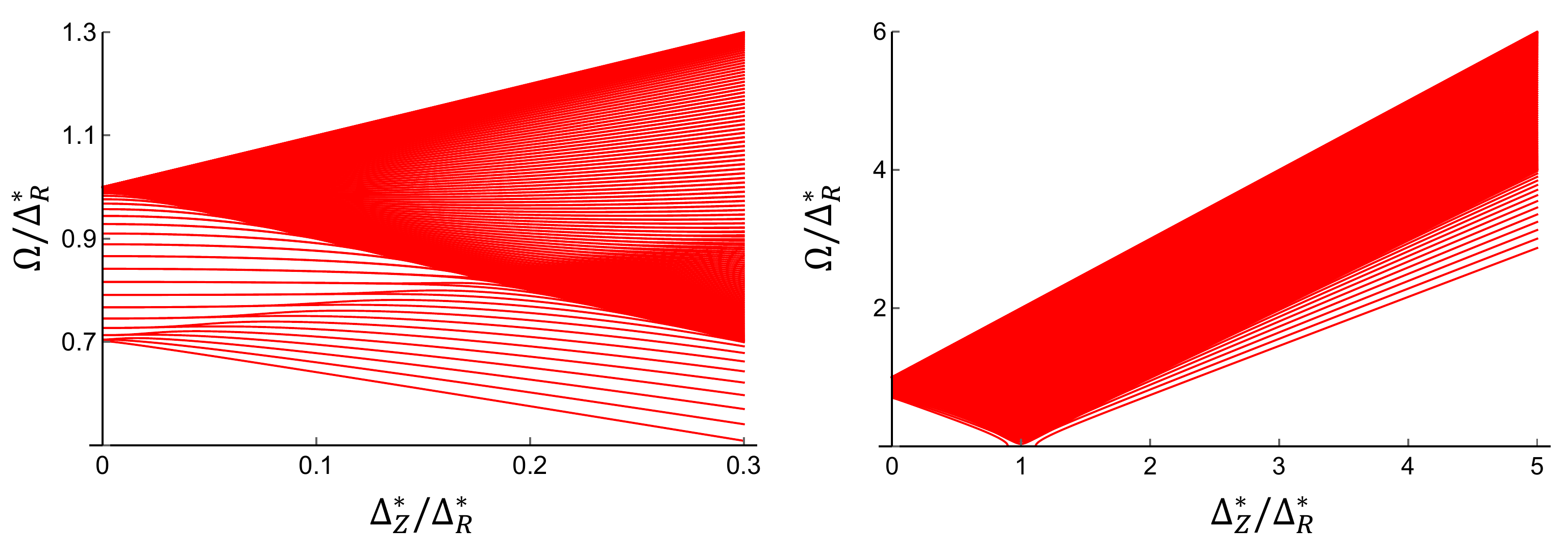}
\caption{\label{fig:beyond_swave} Collective modes of a FL with Rashba SOC (left) and in the presence of the in-plane magnetic field (right) for a model form of the Landau function: $F_m^a = F_0^a \exp(-m^2/m_0^2)$ with $F_0^a=-0.3$ and $m_0 = 10$. Reprinted with permission from Ref.~\onlinecite{kumar_lattice:2017}. Copyright 2017 by the American Physical Society.}
\end{figure*}
If the Landau function contains a large or infinite numbers of harmonics, one has to resort to numerical diagonalization of Eq.~(\ref{TB}).
Figure \ref{fig:beyond_swave} shows the result of such diagonalization for a model form of the  Landau function, given by
$F_m^a = F_0^a \exp(-m^2/m_0^2)$. In this case, the modes are very densely spaced and form a quasi-continuum.
 
 \section{Collective spin modes in Dirac systems}
 \label{sec:Dirac}
\subsection{Graphene with proximity-induced spin-orbit coupling}
\label{sec:GR}
The case of graphene with proximity-induced SOC differs from the one considered in the previous section in two important aspects. First, at the single-particle level, the Hamiltonian \eq{ham0} contains the valley-Zeeman (VZ) term, in addition to the Rashba term.  Second, at the many-body level, the Landau function contains exchange interaction between the valleys. 
In this section, we ignore the asymmetry gap as it has no interesting consequences for the spectrum of the  collective mode. Furthermore, we will neglect a process in which electrons are swapped between the valleys (see Fig.~\ref{scatt}$d$). Then the Landau function becomes isotropic not only in the spin but also in valley subspaces:  
 \bwt
 \bea
\label{FL1}
N_F^*\hat  f(\bk,\bk')
&=&  
F^s(\vartheta) +
\bs
 \cdot\bs'
 F^a(\vartheta)
+
\btau
\cdot \btau'
G^{a}(\vartheta)+ (\bs
\cdot \bs'
)( \btau
\cdot \btau')
H(\vartheta).
\eea
\ewt
Since we are interested in collective modes with frequencies $\ll E_F$, the valence band can be projected out by using L{\"o}wdin method,\cite{lowdin:1951} and the effective single-particle Hamiltonian for the conduction band becomes\cite{Kumar_TMD:2021}
\bea
\label{HR_cc}
\hat H_{c}&=&v_F k + \frac{\lr}{2} (\hat\bk\times \bs)\cdot\hat{\bf z}+\frac{\lz}{2}\tau_z\hat \sigma_z.
\eea
Accordingly, the change in the quasiparticle energy due to SOC is now given by
 \footnote{ Note that in $\tau_z$ is just $\pm1$ in Eq. (\ref{HR_cc}, but it is a matrix  $\hat\tau_z$ in Eq. (\ref{eso}).} 
\bea
\hat{\ve}_{\text{SO}}(\bk)&=&\frac{\lr^*}{2} 
(\hat\bk\times \bs)\cdot\hat{\bf z}
 + \frac{\lz^*}{2}\hat{\tau}_z \hat{\sigma}_z,\nn\\
 &=&\frac{\lr^*}{2}\hat\Sigma_z-\frac{\lz^*}{2}\hat{\tau}_z \hat{\Sigma}_x,\label{eso}
\eea
while the non-equilibrium part of the occupation number can be written as
\bea
\delta \hat n(\bk,t)&=&n_F'\left[\bu(\bk, t) \cdot\bS(\bk) + {\bf w}(\bk, t) \cdot \btau\right.\nn\\
&&\left.  + M_{\alpha\beta}(\bk, t) \hat{\tau}_\alpha \hat{\Sigma}_\beta(\bk)\right],\label{dnsv}
\eea
where the rotated set of Pauli matrices is defined by \Eq{Sigma}.Vector $\bu$ and tensor $M_{\alpha\beta}$  describe oscillations of the uniform 
 \beq
\boldsymbol{\mathcal{S}}=-
\frac{g^*\mu_B}{8} \int \frac{d^{2}k}{(2\pi)^2}  \text{Tr}\left[ \delta \hat n(\bk,t)\, 
\bs\right]
\eeq
and valley-staggered magnetization
\bea
\boldsymbol{\mathcal{M}}
=-
\frac{g^*\mu_B}{8} \int \frac{d^{2}k}{(2\pi)^2}  \text{Tr}\left[ \delta \hat n(\bk,t)\,\hat\tau_z 
\bs\right]
,\label{vsmagn}
\eea
respectively.
The vector ${\bf w}$ describes 
valley polarization, which is decoupled from 
the spin sector and will not be considered below. 

Collective modes of our system correspond to coupled oscillations of the uniform and valley-staggered magnetizations. In the absence of VZ such oscillations are decoupled. The spectrum of collective excitations consists of a doublet of in-plane modes and a single out-of plane mode in each of the sectors, to a total of six modes. VZ SOC mixes the $\boldsymbol{\mathcal{S}}$ and $\boldsymbol{\mathcal{M}}$ sectors. Solving the coupled system of equations of motion, one obtains the following expressions for the  frequencies of the in-plane modes\cite{Kumar_TMD:2021}
\bse
\beq
\label{R+Z}
\begin{split}
\Omega_{||,\pm}^2 =&
\lr^{*2}\bigg( \frac{ff_++hh_+}{2} \bigg) + \lz^{*2} \bigg( f_+h_++f_-h_- \bigg) 
\pm\Omega_{0}^2,
\end{split}
\eeq
where 
\bwt
\beq
\label{O0}
\begin{split}
\Omega_{0}^2 =& \left[\lr^{*4}\bigg( \frac{ff_+-hh_+}{2}\bigg)^2
+\lz^{*4} (f_-h_++h_-f_+)^2+\lr^{*2} \lz^{*2}(ff_-+hh_-)(f_-h_++h_-f_+)\right]^{1/2},
\end{split}
\eeq
\ewt
\ese
 \bea 
f&=&1+F^a_1,\;f_+=1+\frac{F^a_0+F^a_2}{2},\; f_-= \frac{F^a_0-F^a_2}{2},\nn\\
 h&=& 1+H_1,\;h_+=1+\frac{H_0+H_2}{2},\;  h_-= \frac{H_0-H_2}{2},\nn\\
\label{fh}
\eea
and $\lr^*=\lr/(1+F_1^a)$ and $\lz^*=\lz/(1+H_0)$  are the renormalized spin-orbit coupling constants. 
The frequencies of the out-of-plane modes are given by
\bea
\Omega_{\perp,+}^2&=&h\left[(h_++h_-)\lr^{*2}+f\lz^{*2}\right]\nn\\
\Omega_{\perp,-}^2&=&f\left[(f_++f_-)\lr^{*2}+h\lz^{*2}\right]
\eea

Finite $q$ and/or Zeeman magnetic field lift the degeneracy of the  in-plane modes. The spectrum of the collective modes in these cases is a subject of the on-going study.\cite{sam:unpub}
\subsection{Dirac surface state of a 3D topological insulator}
\label{sec:helical}
The Dirac state on the surface of a 3D topological insulator, described by the Hamiltonian \eq{HTI}, is characterized by locking of the spin and charge degrees of freedom. Indeed, \Eq{HTI} implies the charge current and spin densities  are related by an operator identity\cite{raghu:2010} 
\bea
\hat{\bf j}=e\vd\bs\times \hat {\bf z}.\label{js}
\eea
Due to this coupling, the system supports a new kind of collective modes: spin-plasmons.\cite{raghu:2010} Although the dispersion of spin-plasmons at small $q$ is similar to that of usual plasmons in 2DEGs ($\omega\propto \sqrt{q}$), the nature of the two modes is quite different because a spin-plasmon corresponds to coupled oscillations of spin and charge densities. For example, the weight of the spin part at small $q$ is much larger than that of the charge part. 

 \begin{figure}
\centering
\includegraphics[scale=0.17]{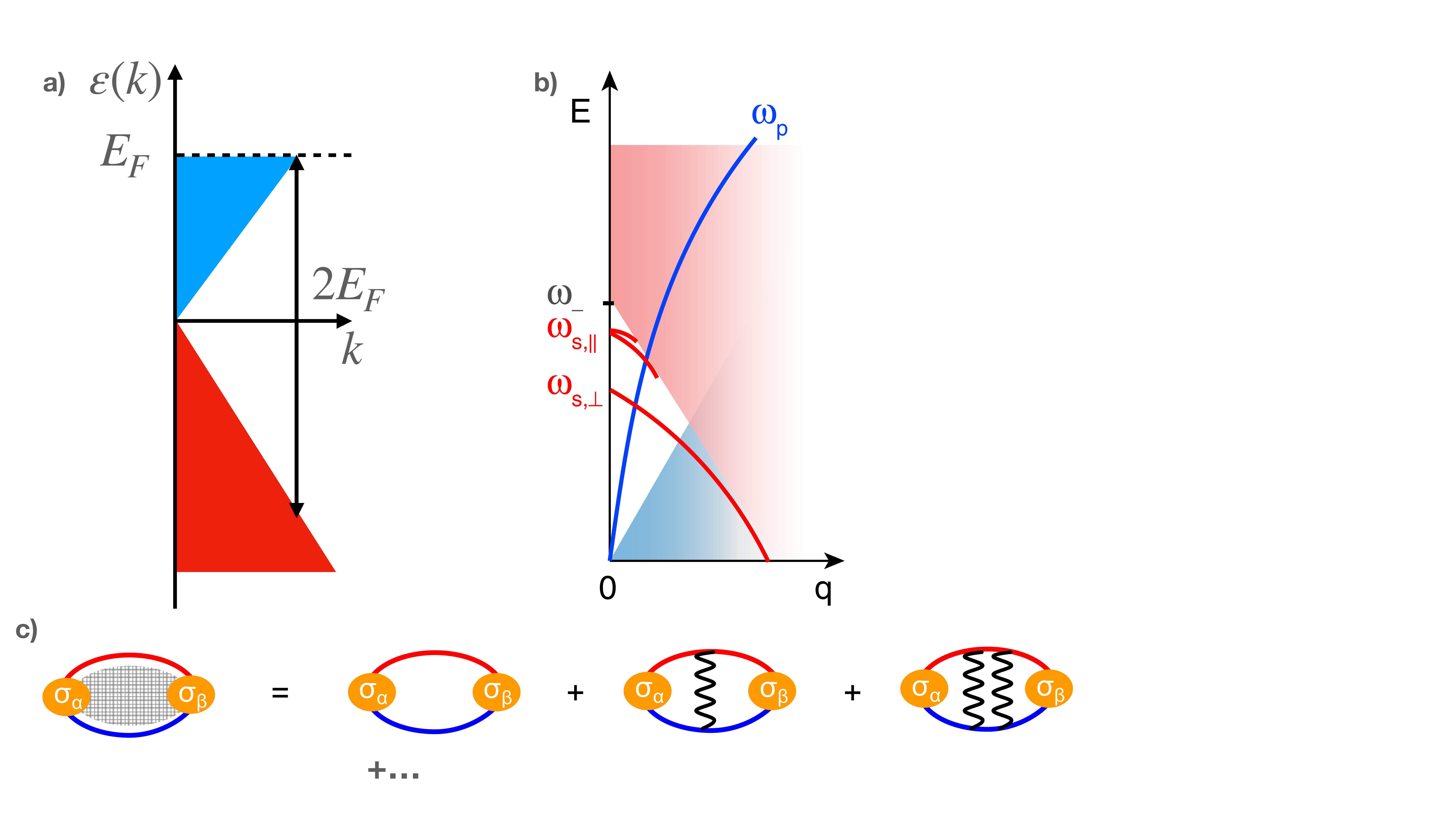}
\caption{\label{fig:helical} a) Spectrum of the Dirac state on the surface of a 3D topological insulator. b) Spectrum of the collective modes of the surface state. $\omega_p$ denotes the spin-plasmon \cite{raghu:2010}, $\omega_{s,||}$ and $\omega_{s,\perp}$ denote the out- and in-plane collective spin modes, and $\omega_-=2E_F$ is the lower boundary of the continuum of spin-flip particle-hole excitations.  c) Ladder diagrams for the spin susceptibility $\chi_{\alpha\beta}$ ($\alpha,\beta=x,y,z$). Panel b) is reprinted with permission from Ref.~\onlinecite{kung:2017}. Copyright 2017 by the American Physical Society.}
\end{figure}

Low-energy excitations of a doped surface state can be described by the effective theory of helical FL, obtained by projecting out the occupied Dirac cone.\cite{lundgren:2015} While the charge sector of such a FL can be described by a few Landau parameters, description of the spin sector runs into difficulties similar to those discussed in Sec.~\ref{sec:noFL}. Namely, the total spin susceptibility contains contribution from high-energy states which cannot be accounted for within the FL theory. Nevertheless, the FS contribution to the spin susceptibility can be expressed through the Landau parameters of a helical FL.\cite{lundgren:2015}

In what follows, we will be interested in a collective mode of the spin-excitonic type observed in Ref.~\onlinecite{kung:2017}; see Sec.~\ref{sec:Bi2Se3} for more details on the experiment. This mode condenses out of the continuum of inter-band particle-hole excitations. In a non-interacting system with Dirac spectrum such a continuum starts at the Pauli threshold $\omega=2E_F$, see Fig.~\ref{fig:helical}$a$, and disperses with $q$ as shown by the shaded region in Fig.~\ref{fig:helical}$b$. The blue shaded region depicts the continuum of gapless intra-band particle-hole excitations. With interactions, one expects to see the spin-plasmon mode (blue curve)
and inter-band collective spin modes (red curves). By the same symmetry arguments as given in Sec.~\ref{sec:RSOC}, there are one out-of-plane mode and two in-plane modes at finite $q$; the latter become degenerate at $q=0$. Inter-band excitations  with energies comparable to $E_F$ cannot be described within the FL theory, and one has to treat electron-electron interaction explicitly. The simplest  method is the ladder approximation, shown diagrammatically in  Fig.~\ref{fig:helical}$c$. For a Hubbard-like interaction and at $q=0$, one obtains the spin susceptibility as \cite{maiti:2014}
\bea 
\label{eq:RPA}
\chi_{\alpha\beta}(\Omega)=
-
\delta_{\alpha\beta}\frac{g^2\mu_B^2}{2}\frac{\Pi_{\alpha\beta}(\Omega)}
{1+\frac{U}{2}
\Pi_{\alpha\beta}(\Omega)},
\eea
where the components of the polarization bubble $\hat\Pi $ are obtained  by analytic
continuation of the corresponding Matsubara expressions
\bwt
\bea\label{eq:Pi}
\Pi_{\alpha\beta}(i\Omega_n)=T\sum_{m} \int\frac{d^2k}{(2\pi)^2} {\rm
	Tr}\left[\hat\sigma_\alpha\hat G_0(\bk,i\varepsilon_m+i\Omega_n)
\hat\sigma_\beta \hat G_0(\bk,i\varepsilon_m)\right].
\eea
\ewt
with $\alpha,\beta=x,y,z$ and 
$G_0(\bk,i\e_m)=
\left(i\e_m-\hat H_{\text{HTI}}\right)^{-1}$
is the Green's function 
of the Hamiltonian \eq{HTI}.

For Dirac spectrum, the momentum integrals for $\Pi_{\alpha\alpha}$  diverge in the ultraviolet and need to be cut off  at some momentum $\Lambda$, chosen from the condition that the cubic term in the dispersion due to hexagonal warping\cite{fu:2009} becomes comparable to the linear one. With such a cutoff, we obtain
\bwt
\bea
\R\Pi_{xx}(\Omega)=\R\Pi_{yy}(\Omega)=\frac 12 \R\Pi_{zz}(\Omega)=-\frac{1}{4\pi \vd}\left(\Lambda+\frac{\Omega}{4\vd}\ln\left\vert\frac{\Omega+2E_F}{\Omega-2E_F}\right\vert\right).
\eea
\ewt
The poles of \Eq{eq:RPA} give the frequencies of the collective modes. At weak coupling, the modes are exponentially close to the boundary of the continuum: 
\bea
\Omega_{||}=2E_F-Ve^{-8/u},\;\Omega_\perp=2E_F-Ve^{-4/u},\label{Hubbard}
\eea
where $u=UE_F/2\pi\vd^2$ is the dimensionless coupling constant and $V=4E_F\exp(\Lambda/2k_F)$.

A reader familiar with the semiconductor literature of the 60s may notice that the derivation presented above is 
very close to that of  ``Mahan excitons'' in degenerate semiconductors.\cite{mahan:1967} Indeed, because of the relation \eq{js} the in-plane components of the spin susceptibility and conductivity are proportional to each other. \cite{raghu:2010} The same analysis, therefore, would predict that the optical conductivity of a helical state should have  a peak at $\Omega=\Omega_{||}$. This is exactly what Mahan obtained by (effectively) resumming the ladder series for the optical conductivity of a degenerate semiconductor; the only difference is that he considered Coulomb rather than Hubbard interaction but this difference is not significant. Indeed, a similar analysis for the Coulomb case shows that the collective modes are exponentially close to the continuum boundary as well, namely\cite{Adamya_thesis}
\bea
\Omega_{||}&=&2E_F\left\{1-\frac{\alpha}{2}\exp\left[-\frac{\pi}{2\ln \alpha^{-1}}\left(\frac{1}{\alpha}- \ln\frac{\Lambda}{k_F} \right)\right]\right\},\nn\\
\Omega_{\perp}&=&2E_F\left\{1-\frac{\alpha}{2}\exp\left[-\frac{\pi}{2\ln \alpha^{-1}}\left(\frac{1}{2\alpha}- \ln\frac{\Lambda}{k_F} \right)\right]\right\},\nn\\
\eea
where $\alpha=e^2/\vd$ is the dimensionless coupling constant of the Coulomb interaction. 

Subsequent analysis showed, however, that the ladder approximation is not adequate for this problem:\cite{gavoret:1969,Pimenov:2017} because electrons and holes interact not in vacuum (as it is the case for an undoped semiconductor) but in the presence of the Fermi sea, the excitonic state is overdamped due to Auger-like electron-hole interactions processes, which start at the indirect threshold of $\Omega=E_F$. Moreover, damping due to electron-electron and electron-hole interactions starts already at $\Omega\ll E_F$, and is still operational at $\Omega\sim E_F$.\cite{Sharma:2021,adamya} 
Both types of damping give rise to the linewidth proportional to $u^2$ (or $\alpha^2$), whereas the binding energy is exponentially small in $1/u$ (or $1/\alpha)$. Therefore, collective modes (and Mahan excitons) must be totally washed out at weak coupling.

Nevertheless, \Eq{eq:RPA} describes very well a spin-resolved collective mode observed by Raman spectroscopy in Bi$_2$Se$_3$, which effectively probes the $zz$-component of the spin susceptibility, see Ref.~\onlinecite{kung:2017} and Sec.~\ref{sec:Bi2Se3}. 
The frequency of the observed mode  is about $E_F$, which implies that the system is not in the weak-coupling regime. On the other hand,
 an order-of magnitude estimate for 
 the linewidth due to all types of electron- and electron-hole interactions is also of order $E_F$ for $\Omega\sim E_F$ and $u\sim 1$ (or $\alpha\sim 1$), so one would expect to see only a very broad peak at best.
Nevertheless, the observed peak is quite sharp: its linewidth is only about $1/20$ of its position. Moreover, the linewidth does not vary with temperature as the latter is raised up to 300 K, which indicates that the broadening is due to impurity scattering. Therefore, damping due to  electron- and electron-hole interactions
is significantly weaker than a naive, order-of-magnitude estimate would suggest. We believe that the reason is purely numerical: as analytical and numerical results show, the linewidth due to both processes is only $5\times 10^{-3}E_F$ for a 2D helical state.\cite{Adamya_thesis,adamya} Therefore, the ladder approximation works much better than it might have been expected to.

\section{Spatial dispersion of collective spin modes}
\label{sec:space}
The dispersion of the transverse $n=0$ Silin mode is quadratic in $q$ for $q\ll \dz^*/v_F^*$:\cite{silin:1958,ma:1968,mineev:2004, mineev:2005} 
 \bea
 \Omega(q)=\dz+
 a_2(\{F^a\})\frac{v_F^{*2}q^2}{\dz},\label{qSilin}
 \eea
 where $a_2(\{F^a\})$ depends on the angular harmonics of the Landau function; explicitly,
 $a_2=(1+F_0^a)^2(1+F_1^a)/(F_0^a-F_1^a)$. Note that $a_2<0$ for $-1<F_0^a<F_1^a<0$ and thus dispersion is downward, as shown in Fig.~\ref{fig:collmodes_B0}.
 A quadratic scaling of the dispersion $q^2$ follows from the invariance of the spin subspace with respect to rotations about the magnetic field, which requires the dispersion to be isotropic, and from analyticity, which requires that an expansion in $q$ starts from the quadratic term. 
 
If only Rashba SOC is present, the group symmetry of the system is $C_{\infty v}$, i.e., the system is invariant with respect to rotations by an arbitrary angle about the normal to the plane. Therefore, as in the Silin's case, the dispersion is isotropic and, by analyticity, starts with a $q^2$ term. For the lowest three modes,\cite{ashrafi:2012,zhang:2013,maiti:2014,maiti:2017} 
\bea
 \Omega_{\alpha}(q)=\Omega_\alpha+
 \tilde a^{\alpha}_2(\{F^a\}) \frac{v_F^{*2}q^2}{\dr},\label{qR}
 \eea
where $\alpha\in\{||x,||y,\perp\}$, the  $q=0$ frequencies are given by Eqs.~\ref{Operp} and Eqs.~\ref{Opar}, and  $\tilde a^{\alpha}_2(\{F^a\})$ some functions of $F_0^a, F_1^a\dots$  
If only Dresselhaus SOC is present, the Hamiltonian \eq{HD} can be transformed back to the Rashba one \eq{HR} by 
by reflecting the spatial coordinates about a mirror plane that contains the (110) axis, upon which $k_x\to k_y$ and $k_y\to k_x$.
Therefore, the spectrum of the collective modes is the same as in \Eq{qR} with $\dr$ being replaced by $\dd$.

If both Rashba and Dresselhaus SOCs are present, the symmetry is lowered to $D_{2d}$, which only allows rotations by $\pi$ and mirror reflections about the two diagonals. The linear term in the dispersion is absent because by symmetry it can only be $\sqrt{q_x^2+q_y^2}$, which is not allowed by analyticity, while a bilinear term is of the   $c_1(q_x^2+q_y^2)+c_2 q_xq_y$ form, where $c_{1,2}$ are constants. 

 An in-plane magnetic field breaks the rotational symmetry. Now, linear-in-$q$ terms in the dispersion are allowed and, indeed, they have been observed experimentally.\cite{perez:2007,baboux:2013,baboux:2015,perez:2016,perez:2017,perez:2019}  The structure of such terms can be deduced just from the symmetries of $C_{\infty v}$ and $D_{2d}$ groups.\cite{baboux:2015,perez:2016,maiti:2017} In both cases, we need to form a scalar out of a polar vector $\bq$ and axial vector ${\bf B}$. In the $C_{\infty v}$ group, this is only possible by forming the Rashba invariant $B_xq_y-B_yq_x=Bq \sin(\phi_\bq-\phi_{{\bf B}})$, were $\phi_\bq$ and $\phi_{{\bf B}}$ are the azimuthal angles of $\bq$ and ${\bf B}$, respectively. Likewise, the only scalar that can be constructed in the $D_{2d}$ group is the Dresselhaus invariant $B_xq_x-B_yq_y=Bq \cos(\phi_\bq+\phi_{{\bf B}})$.
 
 In addition to linear-in-$q$ terms, an in-plane magnetic field also gives rise to the dependence of the mode frequency at $q=0$ on the direction of the magnetic field. However, this effect needs both Rashba and Dresselhaus SOCs to be present. Indeed, since Rashba SOC has continuous rotational symmetry, the direction of the magnetic field is irrelevant in this case, while the case of pure Dresselhaus SOC is reduced back to pure Rashba one. If both types of SOC are present, the $D_{2d}$ symmetry implies that mode frequency depends on the direction of ${\bf B}$ as $\Omega_\alpha(q=0)\propto \sin 2\phi_{{\bf B}}$.\cite{maiti:2017,perez:2017}
 
 To be specific, from this point onwards we will focus on the case when $\dz\gg\dr,\dd\neq 0$, which is relevant for Raman experiments on Cd$_{1-x}$Mn$_{x}$Te quantum wells.\cite{perez:2007,baboux:2013,baboux:2015,perez:2016,perez:2017} In this case, the mode frequency at $q=0$ is given primarily by the (bare) Zeeman energy. The correction due to SOC must be quadratic in both $\dr$ and $\dd$ and symmetric on $\dr\leftrightarrow\dd$, i.e., to be of the form $\dr^2+\dd^2$. The anisotropic term at the $q=0$ exists only of both $\dr$ and $\dd$ are non-zero which, in the limit considered, implies that the corresponding term must be proportional to $\dr\dd$. The linear-in-$q$ terms due to Rashba and Dresselhaus SOC must be proportional to $\dr$ and $\dd$, respectively. Finally, the bilinear-in-$q$ is almost the same as for the Silin mode modulo an anisotropic correction, which arises again only due to the combined effect of Rashba and Dresselhaus SOCs, and thus is also proportional to $\dr\dd$. Combining all the arguments given above, we arrive at the following form of the dispersion\cite{perez:2016,maiti:2017,perez:2017}
 \bea
 \Omega(\bq,{\bf B})= \Omega_0({\bf B})+w(\bq,{\bf B})q+S(\bq)\frac{v_F^{*2}q^2}{\dz},\nn\\
 \eea
 where 
 \bea
 \Omega_0({\bf B})&=&\dz+a_0(\{F^a\})\frac{\dr^2+\dd^2}{\dz}\nn\\
 &+& \tilde a_0(\{F^a\}) \frac{\dr\dd}{\dz}\sin 2\phi_{{\bf B}},\nn\\
 w(\bq,{\bf B})&=&v_F a_1(\{F^a\})\left[\frac{\dr}{\dz}\sin(\phi_\bq-\phi_{{\bf B}})\right.\nn\\
 &&\left.+\frac{\dd}{\dz}\cos(\phi_\bq+\phi_{{\bf B}})\right]\nn\\
 S(\bq)&=& a_2(\{F^a\})+\tilde a_2(\{F^a\})\frac{\dr\dd}{\dz}\sin 2\phi_\bq,\label{qq2}
 \eea
and the coefficient $a_2(\{F^a\})$ is the same as for a pure Silin mode, \Eq{qSilin}. In the equation above we omitted isotropic $q^2$ terms proportional to $\dr^2$ and $\dd^2$. The coefficients $a_0$, $\tilde a_0$, and $a_1$ were calculated in Ref.~\onlinecite{maiti:2017} in the $s$-wave approximation for the Landau function, see Eq.~(20) in there.
 It was argued in Refs.~\onlinecite{perez:2016,perez:2017} that, to linear order in SOC, the dispersion can be obtained by a canonical transformation of the Hamiltonian, which amounts to replacing $\bq$ in the dispersion of the Silin mode, \Eq{qSilin}, by $\bq+\bq_0$, where $\bq_0$ is proportional to $\dr$ and $\dd$. This would imply that the coefficients of the $q$ and $q^2$ terms in \Eq{qq2} are related as $|a_1|=2|a_2|$. However, a microscopic calculation\cite{maiti:2016} shows that such relation is satisfied only in the weak-coupling limit ($|F_0^a|\ll 1$).

\section{Damping of collective spin modes}
\label{sec:damping}

Spin-orbit coupling is the reason for the collective spin modes, described in this paper, to exist. At the same time, however, it couples spin and momentum and thus enables damping of spin excitations via momentum relaxation due to scattering by non-magnetic degrees of freedom: non-magnetic disorder, phonons, etc. 

\begin{figure}
\centering
\includegraphics[width=1\columnwidth]{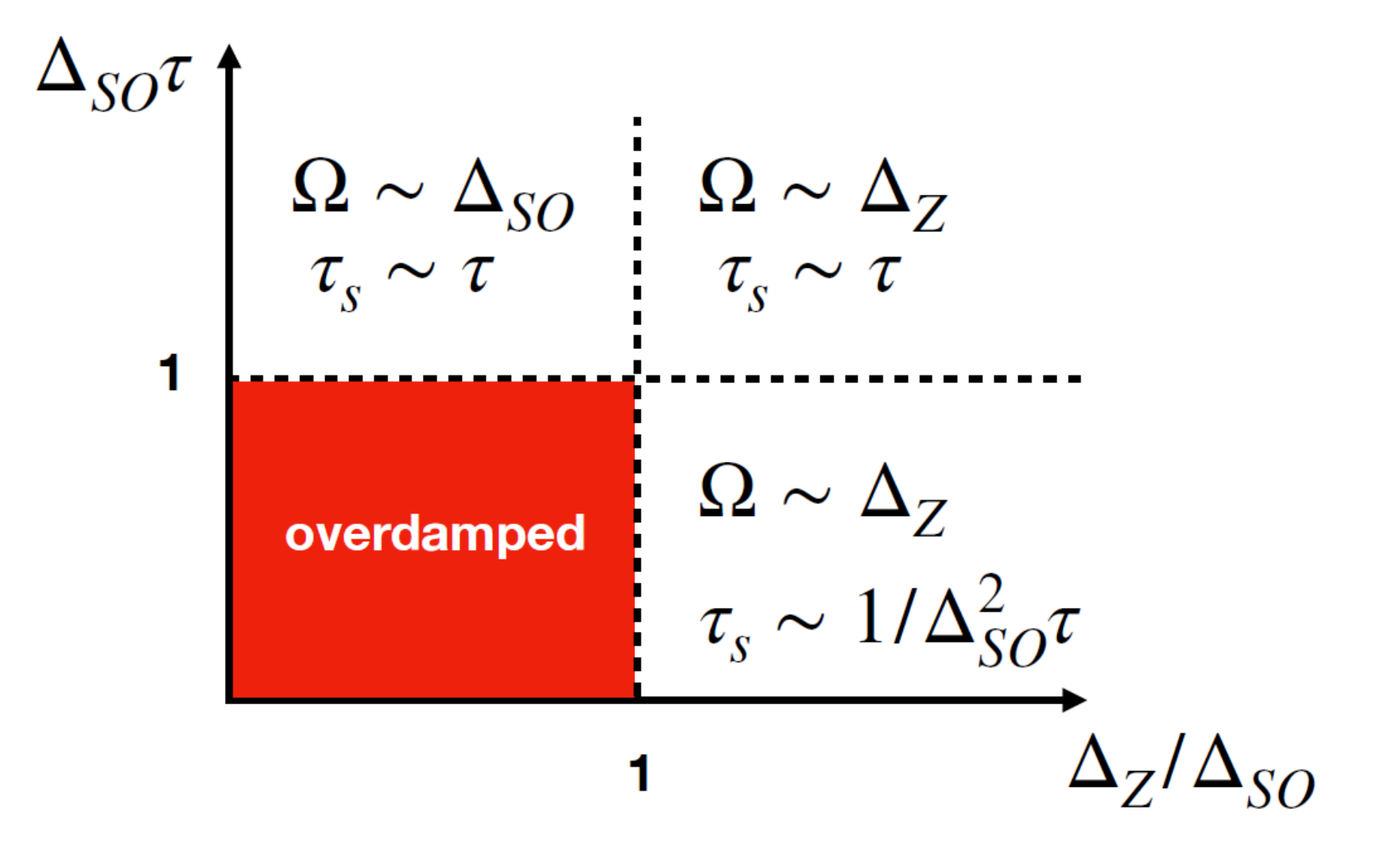}
\caption{\label{fig:DP} Damping of collective spin modes by non-magnetic disorder via D'yakonov-Perel' mechanism. $\Delta_{SO}$ and $\Delta$ are the splittings of the energy spectrum due to SOC and magnetic field, respectively,  $\tau$ is the transport mean free time due to scattering by disorder. In the region outside the red square the spin collective modes are underdamped. The frequency of the collective mode $(\Omega$) and spin relaxation time ($\tau_s$) are indicated for each region.  Within the red square, the modes are overdamped and thus cannot be observed.}
\end{figure}

Non-magnetic disorder in combination with SOC leads to spin relaxation via the Elliott-Yaffet and D'yakonov-Perel' (DP) mechanisms.\cite{dyakonov:book}  The former is present in both centro- and non-centrosymmetric systems, while the latter is specific for non-centrosymmetric systems, considered in this article, and in this case the DP mechanism is usually the dominant one.  If $\tau$ is a characteristic time of momentum relaxation by disorder (to be defined more precisely later) and $\Delta_{\text{SO}}$ is the energy splitting due to SOC, then the spin dynamics is ballistic with the DP spin relaxation time $\tau_s\sim\tau$ for  $\Delta _{\text{SO}}\tau\gg 1$, and diffusive with the DP spin relaxation time $\tau_s\sim 1/\Delta^2_{\text{SO}}\tau$ for $\Delta_{\text{SO}}\tau\ll 1$.  On the other hand, the frequency of the collective mode is set by the largest of two energy scales: $\Delta_{\text{SO}}$ and the Zeeman energy $\dz$.  The spin collective mode can be resolved only when it is underdamped, which corresponds to the region outside the red square in Fig.~\ref{fig:DP}. This requires clean samples and strong SOC/magnetic fields.

 One concrete example is a single-valley 2D FL with disorder and Rashba SOC. In this case, the width of the chiral spin resonance at $q=0$ is controlled by two relaxation times, renormalized by the FL interaction. For $\dr^*\tau\gg 1$, the width of the resonance is\cite{shekhter:2005}
\bea
\Gamma_{\text{i}}=\frac{1}{2x_1\tau_1}+\frac{1}{2x_2(1+x_2/x_0)\tau_2},
\eea
where $x_m=1/(1+F^a_m)$,
\bea
\frac{1}{\tau_{n}}=\pi N_F^* n_{\text{i}} \int \frac{d\vartheta}{2\pi} (1-\cos n\vartheta)\left\vert u\left(2k_F\sin \frac{\vartheta}{2}\right)\right\vert^2,\nn\\
\eea
$n_{\text{i}}$ is the number density of impurities, and $u(Q)$ is the Fourier transform of the single-impurity potential. Note that $\tau_1$ is the transport time that enters the mobility.

\begin{figure}
\centering
\includegraphics[width=1\columnwidth]{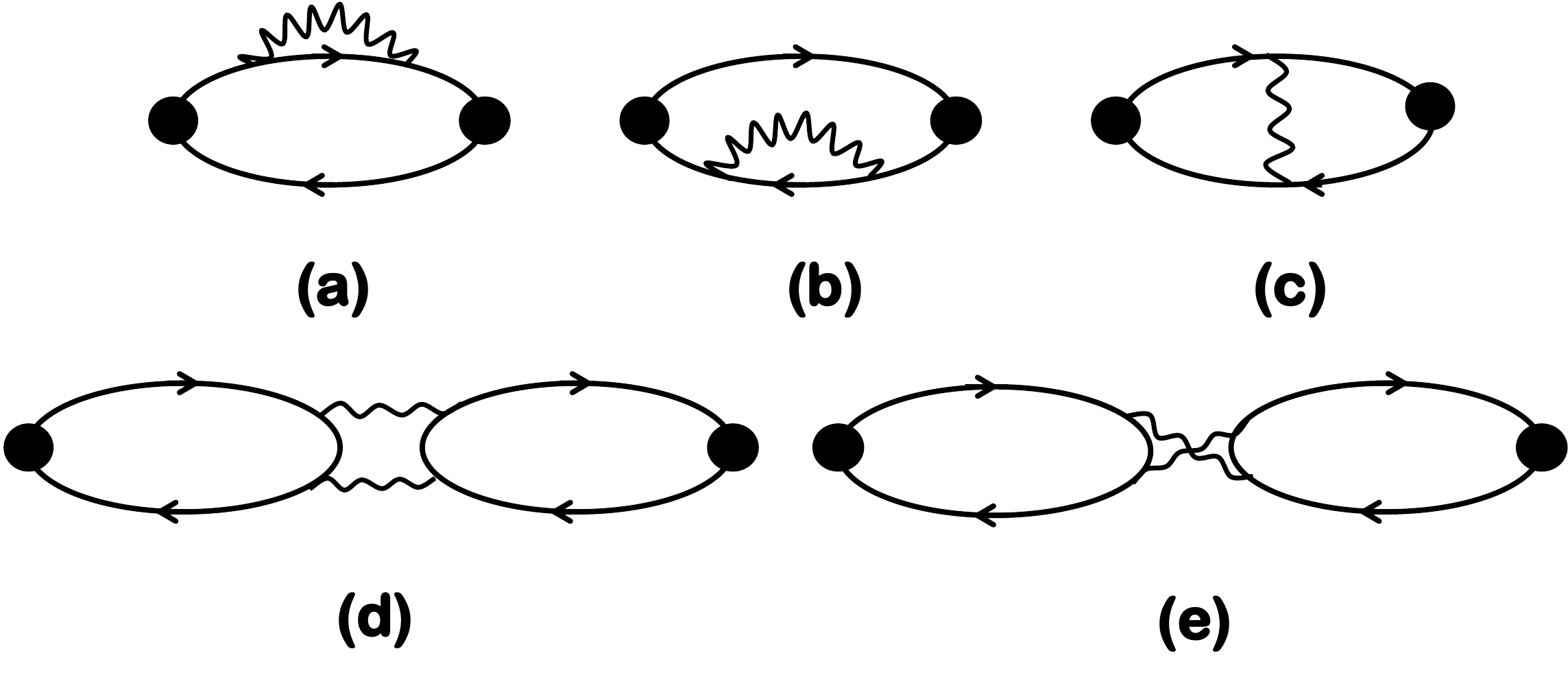}
\caption{\label{fig:damping} Diagrams contributing to damping of the collective modes by dynamically screened Coulomb interaction (wavy line). The filled circles denote the corresponding vertices. For the case of spin collective modes, the vertices are Pauli matrices.}
\end{figure}

Another source of damping is electron-electron interaction. Because the collective modes lie outside the particle-hole continuum, they are not affected by Landau damping, which involves excitation of a single particle-hole pair. However, excitations involving multiple pairs are possible due to the residual interaction between FL quasiparticles, and such excitations give finite width even to modes outside the continuum. For example, plasmons, \cite{dubois:1969,mishchenko:2004,pustilnik:2006,imambekov:2012,principi:2013,Sharma:2021} Silin
mode in a partially spin-polarized FL,\cite{ma:1968,mineev:2004, mineev:2005} and magnons in a ferromagnetic FL\cite{kondratenko:1976,mineev:2004, mineev:2005} are all 
damped via this mechanism. Figure \ref{fig:damping} shows diagrams that contribute to damping to lowest order in a dynamically screened Coulomb interaction.\footnote{Although the two last (Aslamazov-Larkin) diagrams appear to be higher order in the Coulomb interaction than the first three, they actually contribute to the same order in the dimensionless coupling constant $e^2/\hbar v_F$.} 

Out of the examples listed above, the Silin mode is the closest one to collective spin modes due to SOC. However, there is an important difference. 
Namely, conservation of the spin component  along the field 
ensures that the Silin mode is not damped at
$q\rightarrow0$.\cite{kondratenko:1965} On the contrary, spin is not a conserved quantity in a FL with SOC. Therefore, collective modes in this case are damped even at $q=0$. Evaluation of diagrams in Fig.~\ref{fig:damping} leads to an intuitively clear result: at $T=0$, the width of the resonance is given by\cite{maiti:2015}
\bea
\Gamma_{\text{ee}}\sim \frac{\Delta_{\text{SO}}^2}{E_F}\lambda^2\ln \lambda^{-1},
\eea
 where $\lambda=e^2/v_F$ is the dimensionless coupling constant of the Coulomb interaction. The quadratic dependence on $\Delta_{\text{SO}}$ is an expected scaling of a relaxation rate in a FL.\footnote{A single-particle relaxation rate in a 2D FL has an additional logarithmic factor, but it is canceled out between the diagrams for the spin relaxation rate, as guaranteed by the gauge invariance.} 
 
 Modes with finite $q$ are damped by the electron-electron interaction even in the absence of SOC. The same arguments of rotational invariance and analyticity that we used in Sec.~\ref{sec:space} to determine the $q$-dependence of the dispersion, can be applied to the damping rate. 
 Namely, in the presence of rotational symmetry, the damping rate  is proportional to $q^2$.\cite{ma:1968,mineev:2004} If both types of SOC and an in-plane magnetic field are present, the damping rate contains a linear-in-$q$ term, whose structure is the same as of the linear term in the dispersion.\cite{perez:2017,perez:2019}
 
%

\section{Experiment:\\observations and predictions}
\label{sec:exp}
 \subsection{Collective spin waves in $\text{Cd}_{1-x}\text{Mn}_x\text{Te}$  quantum wells}
\label{sec:finiteq}
The dispersion of collective spin waves in  Cd$_{1-x}$Mn$_x$Te  quantum wells was measured  in a series of detailed Raman experiments.\cite{perez:2007,baboux:2013,baboux:2015,perez:2016,perez:2017} This system has both Rashba and Dresselhaus SOCs and thus should have chiral spin waves even in the zero magnetic field. However, the corresponding energy scales are below the resolution of Raman spectroscopy, and one has to apply an in-plane magnetic field to increase the energies of spin-flip excitations. Even a moderate field of 2 T leads to a significant spin polarization due the exchange interaction between magnetic moments of Mn dopants and conduction electron spins, and the effective Zeeman energy is larger than the spin-orbit one. To get some sense of the numbers, for the reported values of $\alpha^*\approx 1.8$\,meV$\cdot$\AA\,and $\beta^*\approx 3.8$\,meV$\cdot$\AA\, (Ref.~\onlinecite{perez:2016}),\footnote{We assigned stars to $\alpha$ and $\beta$ because the experiment measures only the renormalized values of the spin-orbit parameters.} the combined spin-orbit splitting is $\Delta^*_{\text{SO}}=\sqrt{\dr^{*2}+\dd^{*2}}\approx 0.1$\,meV at $n=2.7\times 10^{11}\text{cm}^{-2}$, while the measured frequency of the collective mode varies from $\sim 0.4$\, meV at $q=0$ in samples with lower Mn fraction($x=0.013$, Ref.~\onlinecite{perez:2016} ) to $\sim 3$\,mev in samples with higher Mn fraction ($x=0.8$, Ref.~\onlinecite{perez:2007}). Therefore,  the effective Zeeman splitting is 4-30 times larger than the spin-orbit one, and the experimental situation corresponds to the right panel of Fig.~\ref{fig:R+B_s_wave} for $\dz^*>\dr^*$ (with $\dr^*$ replaced by $\Delta^*_{\text{SO}}$), i.e., there is  a single collective mode, which evolves into a pure Silin mode in the limit of $\dz^*\to\infty$.   At lower Mn fractions, however, the effects of SOC on the dispersion and damping of this mode are quite pronounced. 

\begin{figure}
\centering
\includegraphics[width=1\columnwidth]{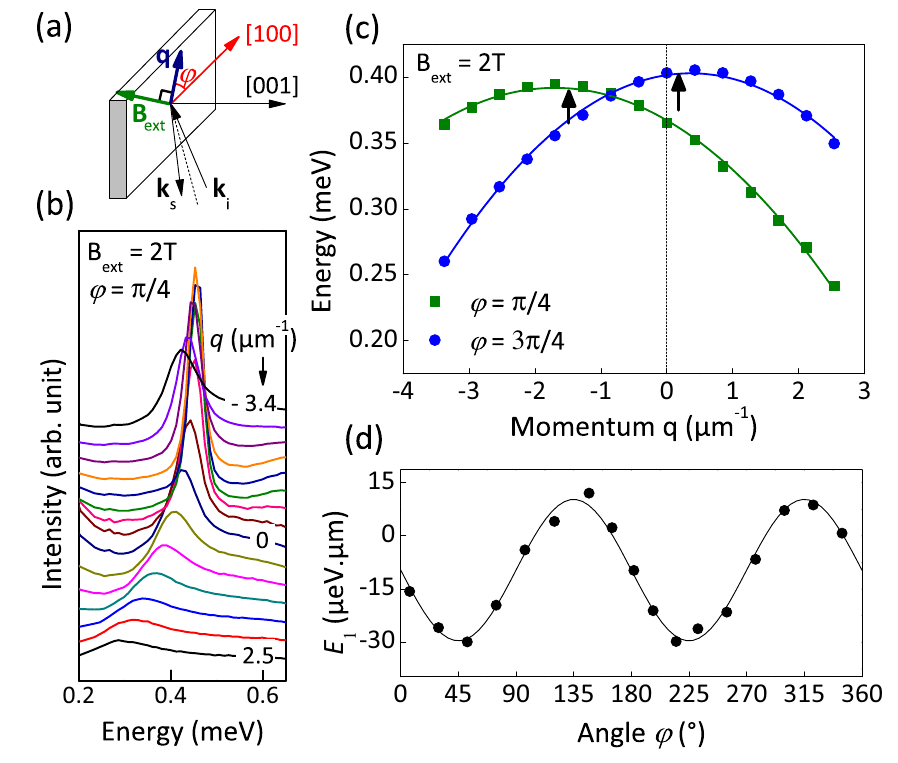}
\includegraphics[scale=0.4]{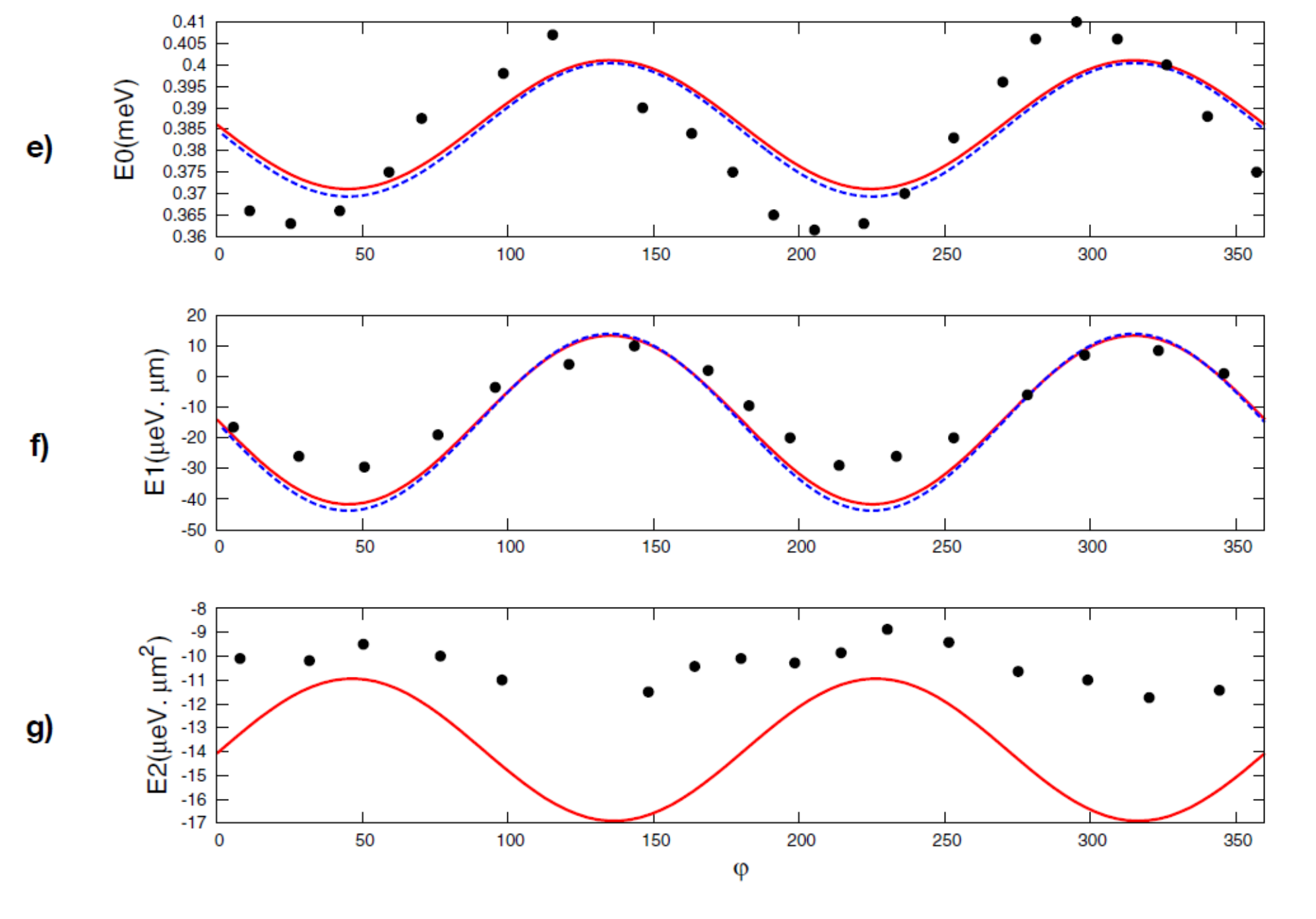}
\caption{\label{fig:perez_2015} Experimental results for collective spin waves in $Cd_{1-x}$Mn$_x$Te quantum wells. a) Geometry of the Raman experiment. b) Raman signal. c) $q$-dependences of the spin-wave frequency. d) Angular dependence of the spin-wave velocity [denoted by $w({\bf B})$ in the main text]. Panels $e$-$g$: angular dependence of the mode frequency at $q=0$ [denoted by $\Omega_0({\bf B})$ in the main text], and spin-wave stiffness 
[denoted by $S({\bf B})$ in the main text], respectively. Solid and dashed lines represent the numerical and analytical results of Ref.~\onlinecite{perez:2017}. Panels $a$-$d$ and $e$-$f$ are reproduced with permission from Ref.~\onlinecite{baboux:2015} and \onlinecite{perez:2017}, respectively. Copyright 2015, 2017 of the American Physical Society.}
\end{figure}

Panels $a$-$d$ in Fig.~\ref{fig:perez_2015} show experimental data for a [001] Cd$_{87}$Mn$_{0.13}$Te quantum well, reproduced from Refs.~\onlinecite{baboux:2015} and Refs.~\cite{perez:2017}.  In the experiment, the in-plane magnetic field and vector $\bq$ were kept at $90^{\circ}$ to each other, while the pair of vectors was rotated by angle $\varphi$ measured from the [100] direction, as shown in panel $a$. The Raman signal in panel $b$ exhibits a well-resolved peak which disperses with $q$. Interestingly, the dispersion is not purely quadratic but has a sizable linear term, which is revealed by flipping the direction of $\bq$, as shown in panel $c$. Panels $e$-$g$, reproduced from Ref.~\onlinecite{perez:2017}, show the angular dependence of the mode frequency ($E_0$), spin-wave velocity ($E_1$), and spin-wave stiffness ($E_2$). 

Now we compare the experimental results with the theory presented in Sec.~\ref{sec:space}. Under the condition $\phi_{{\bf B}}-\phi_{\bq}=\pm \pi/2$, parameters of the dispersion in \Eq{qq2} are reduced to
 \bea
 \Omega_0({\bf B})&=&\dz+a_0(\{F^a\})\frac{\dr^2+\dd^2}{\dz}\nn\\
 &+& \tilde a_0(\{F^a\}) \frac{\dr\dd}{\dz}\sin 2\phi_{{\bf B}},\nn\\
 w({\bf B})&=&\pm v_F a_1(\{F^a\})
 \frac{\dr-\dd\sin 2\phi_{{\bf B}}}{\dz}\nn\\
 S({\bf B})&=& a_2(\{F^a\})-\tilde a_2(\{F^a\})\frac{\dr\dd}{\dz}\sin 2\phi_{{\bf B}}.\label{qq2b}
 \eea

\begin{figure}
\includegraphics[width=1\columnwidth]{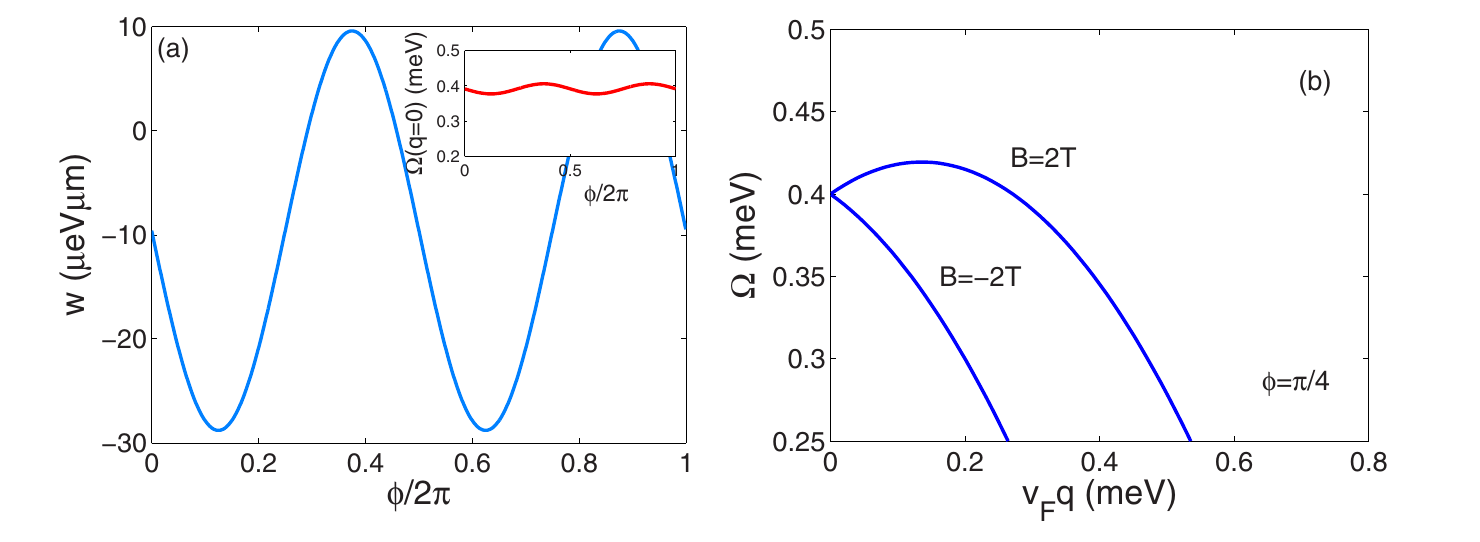}
\caption{\label{fig:Raman} Theoretical results for the parameters of the spin-wave dispersion presented in Eq.~(\ref{qq2b}). Reprinted with permission from Ref.~\onlinecite{maiti:2017}. Copyright 2017 by the American Physical Society.}
\end{figure}

Figure \ref{fig:Raman} shows the theoretical results for $\Omega_0({\bf B})$  (inset in panel $a$), $w({\bf B})$ (panel $a$), and the $q$- dependence of the dispersion for two opposite orientations of the magnetic field, at $\pi/4$ and $-\pi/4$ (panel $b$). The value of $\dz=0.4$\,meV at $B=2$\,T was taken from the experiment, while the Rashba and Dresselhaus coupling constants, and $F_0^a$ were used as fitting parameters. The fitted values of $\alpha$ and $\beta$ (1.9 mev\,\AA\, and 3.8 mev\,\AA) are very close to that reported in Ref.~\onlinecite{perez:2016} ($1.83\pm 0.08$\,meV \AA\, and $3.79\pm 0.11$\,meV \AA, respectively), while the fitted value $F_0^a=-0.41$ is quite close to the Hartree-Fock estimate $F_0^a=-0.3$ for a CdTe quantum well with $n=2.7\times 10^{11}\text{cm}^{-2}$.\cite{maiti:2017} We see that the spin-wave velocity ($w$) is indeed $\pi$-periodic and very close in magnitude the experimental result in panel $f$ of Fig.~\ref{fig:perez_2015}. The $\pi$-modulation of $\Omega_0({\bf B})$, shown in the inset of panel $a$, 
is much smaller than that of $w$ because the former effect is second order in SOC.\cite{maiti:2017,perez:2017} This is consistent with panel $e$ of Fig.~\ref{fig:perez_2015}. The linear term in the dispersion is evident from panel $b$ of  Fig.~\ref{fig:Raman}  and consistent with panel $c$ of Fig.~\ref{fig:perez_2015}. Finally, the experiment also observes small $\pi$-modulation of the spin stiffness (panel $g$), which is consistent with this effect being also second order in SOC.

\subsection{Collective spin mode on the surface of B$\text{i}_2$S${\text{e}_3}$}
\label{sec:Bi2Se3}
\begin{figure*}
\includegraphics[scale=0.28]{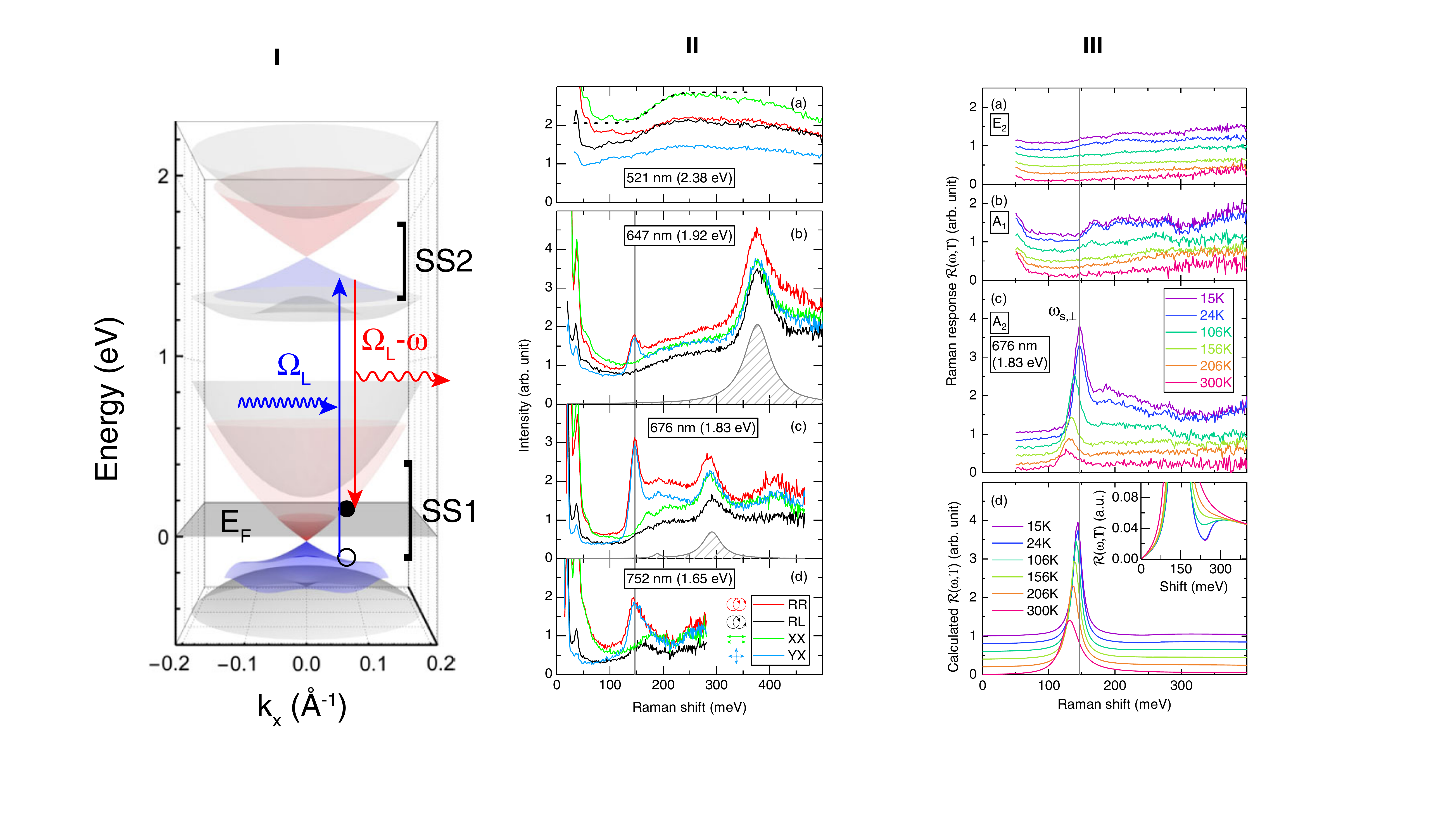}
\caption{\label{fig:kung2017} I) Bandstructure of Bi$_2$Se$_3$ around the Brillouin zone center. The two groups of Dirac States (SS1 and SS2, respectively) are separated by $\Delta_S\approx 1.8$\,eV. The frequency of incident photons $\Omega_L$ is tuned to a resonance transition between the SS1 and SS2 states.
II) Raman scattering data for different polarization geometries of the incoming and scattered photons, and 
different incident photon energies. The collective mode at the Raman shift of $150~$meV is resonantly enhanced by the 1.83 eV photon. III (a)-(c): Raman scattering data decomposed into different channels, corresponding to $E_2$, $A_1$, and $A_2$ irreducible representations of the $C_{6v}$ group, respectively.  
III (d): Theoretical results for the Raman response within the ladder approximation. Reprinted with permission from Ref.~\onlinecite{kung:2017}. Copyright 2017 by the American Physical Society.}
\end{figure*}

Figure \ref{fig:kung2017} summarizes the results of polarization-resolved Raman spectroscopy of the surface state of a topological insulator Bi$_2$Se$_3$.\cite{kung:2017} To enhance the signal from the surface states, the frequency of incident light was tuned to the transition between two surface Dirac points: near the Fermi energy (SS1) and about $\Delta_S=1.8$\, eV above it (SS2), see panel I in Fig.~\ref{fig:kung2017}.   The Raman response shows a well-resolved peak at $\approx 150$\,meV, while the Pauli threshold for this sample is at $2E_F\approx 260$\,meV. The peak is most pronounced for the incident frequency of $1.83$\,eV (panel II$c$), which is the closest to $\Delta_S$.   This proves that the signal is indeed due to the surface rather than the bulk states.  (Broader peaks at higher frequency, interpreted as the result of excitonic photoluminescence,\cite{Kung:2019} were eliminated by subtracting the hatched areas from the data.) Furthermore, polarization-resolved experiments reveal a magnetic nature of the $150$\,meV excitation. As shown in panel II$d$, the signal is much stronger in the RR channel, when right-polarized photons are scattered into right-polarized ones, i.e,  when the angular momenta of the incident and scattered photons differ by $2\hbar$,  than in the RL channel, when right-polarized photons are scattered into left-polarized ones, i.e.,  without a change in the angular momentum. For linearly polarized light, the signal is the strongest in the cross-polarization channel (XY), when the polarization axis is rotated by $\pi/2$. Decomposing the signal into components corresponding to irreducible representations of the $C_{6v}$ group, one finds that the $150$\,meV excitation belongs primarily to the $A_2$ representation, which is the pseudovector representation of $C_{6v}$, see panels III $a$-$c$. Noticeably, the excitation is very robust--it is observed up to 300\,K, see panel III$c$.

Given the findings summarized above, one is prompted to interpret the $150$\,meV excitation as the $q=0$ collective spin mode discussed in Sec.~\ref{sec:helical}, namely, as the out-of-plane mode with frequency $\Omega_\perp$, because in the experiment both the incident and scattered beams were along the normal to the surface. This interpretation is confirmed by the theoretical analysis, which shows that the Raman intensity is proportional to the $zz$-component of the spin susceptibility\cite{kung:2017}
\bea
{\mathcal R}(\Omega,T)\propto \chi_{zz}(\omega,T)\frac{\gamma_{\text r}}{\left(\Omega_L-\Delta_{S}\right)^2+\gamma_{\text{r}}^2}, 
\eea
where $\Omega_L$ is the frequency of incident light
and $\gamma_{\text{r}}$ is the linewidth of the resonance.
$\chi_{zz}(\omega,T)$ was calculated within the ladder approximation for a realistic spectrum of the surface state and at finite temperature, and without using the weak-coupling assumption employed in Sec.~\ref{sec:helical}. [For reasons discussed in Sec.~\ref{sec:helical}, the ladder approximation works better than it might have been expected to, thanks to (accidental) numerical smallness of damping due to electron- and electron-hole interactions.] Since the linewidth of the observed peak is independent of temperature, the primary source of damping must be due to disorder. To mimic the effect of damping, the calculated $\chi_{zz}(\omega,T)$ was artificially broadened to produce the observed linewidth of $\approx 8$\,meV. In addition to the linewidth, the coupling constant of a Hubbard-like interaction was treated as a fitting parameter. The theoretical results, presented in panel III$d$, reproduce very well not only the profile of the peak as a function of frequency but also its temperature dependence. In particular, the theory reproduces the pronounced decrease of the resonance frequency with increasing temperature. This happens because the continuum broadens as $T$ increases, which  pushes the resonance peak down to lower frequency. The inset of panel III$d$ shows a zoom on the interval between the resonance peak and continuum boundary, which is supposed to be at $2E_F\approx 260$\,meV. We see, however, that the continuum is barely discernible because most of its spectral weight is transferred to the collective mode. The best fit was obtained for the Hubbard coupling $u\approx 0.6$. This is consistent with the 
estimate for the screened Coulomb interaction between electrons on the surface of Bi$_2$Se$_3$. 

In summary, the results discussed above present strong evidence for a new type of the collective spin mode, arising from the combined effect of SOC and Coulomb interaction. Further experiments measuring the spatial dispersion of this mode would be highly desirable.
 
\subsection{Predictions for future experiments}
\label{sec:ESR_EDSR}
\subsubsection{ESR and EDSR in 2DEGs}
The rate of absorption of electromagnetic wave at normal incidence by a 2DEG with Rashba and/or Dresselhaus SOC  can be written as
\bea
Q=\frac 12\left[\mu_B^2\Omega\chi''_{||}(\Omega)+\sigma'(\Omega)\right]E_{\text{em}}^2,\label{abs}
\eea 
where $E_{\text{em}}$ is the amplitude of the electric field of the wave, $\sigma(\Omega)$ is the conductivity of a 2DEG and $\chi_{||}(\Omega)$ is the in-plane component of its spin susceptibility. The first term  represents absorption due to ESR, which can be observed even in the absence of SOC. However, its magnitude is proportional to $1/c^{2}$ (in the Gaussian unit system) and thus small. In 2D systems  instead of ESR one typically measures  electrically-detected spin resonance, 
observed as a peak in the longitudinal resistivity under microwave radiation  in the regime of integer Hall effect. \cite{stein:1983,graeff:1999,bowers:2000,bowers:2003,sichau:2019} 

The second term in \Eq{abs} 
represents absorption due to EDSR.\cite{rashba:1991,rashbaefros2003,efros2006}
 Its origin is an effective magnetic field acting on electron spins due to SOC and with magnitude proportional to $k$. The driving electric field (either from a {\em dc} current or electromagnetic  wave) gives rise to a flow of electrons with a non-zero drift velocity, and hence the electron system as a whole experiences an effective magnetic field due to SOC. 
This effect gives rise to a range of spectacular phenomena, e.g., a strong enhancement  of microwave absorption in a geometry when the driving electric field is in the plane of a 2DEG~\cite{schulte:2005} and a shift of the spin resonance frequency by {\em dc} current.~\cite{wilamowski:2007,wilamowski:2008}

Collective spin modes described in the previous parts of the paper correspond to oscillations of the electron magnetization even in the absence of the external magnetic field. Therefore, they should be detectable both via ESR and EDSR which, in contrast to the conventional setup, should be present even in zero magnetic field. The structure of the signal can be understood qualitatively from Fig.~\ref{fig:R+B_s_wave}, where one just has to replace $\dr\to\sqrt{\dr^2+\dd^2}$ to account for Dresselhaus SOC. In the absence of the magnetic field ($\dz^*=0$), the signal consists of two peaks, at frequencies $\Omega_\perp$ and $\Omega_{||}$. At finite field, the peak at $\Omega_{||}$ splits into two. Upon further increase of the field, the $\Omega_{||}$ peaks merge with the continuum and die out, while the $\Omega_\perp$ peak continues to be present all the way till the gap closing point ($\dz^*=\sqrt{\dr^2+\dd^2}$), and then emerges again at fields above
this point. 

To estimate the relative strength of the ESR and EDSR signals, we note that the conductivity in \Eq{abs} is the sum of the Drude and spin-orbit parts: $\sigma'(\Omega)=\sigma'_{\text{D}}(\Omega)+\sigma'_{\text{SO}}(\Omega)$.
Since for the Hamiltonians \eq{HR} and \eq{HD} the electric current is proportional to magnetization, the spin-orbit part of the conductivity and spin susceptibility are related by
$\sigma'_{\text{SO}}(\Omega)\sim e^2\max\{\alpha^2,\beta^2\} \chi''(\Omega)/\Omega$. Provided that the Drude part at the resonance frequency $\Omega_{\text{r}}$ can be neglected, the ratio of the EDSR to ESR signals can be estimated as\cite{shekhter:2005}
\beq
\begin{split}
\frac{Q_{\text{EDSR}}}{Q_{\text{ESR}}} &\sim \left(\frac{\max\{\alpha,\beta\}mc}{\Omega_{\text{r}}}\right)^2 \\
&\sim \left(\frac{\lambda_F}{\lambda_C}\right)^2\left\{
\begin{array}{ccc}
1,\;\text{for}\;\dz\ll \max\{\dr,\dd\},\\
\left(\frac{\max\{\dr,\dd\}}{\dz}\right)^2,
\end{array}
\right.
\end{split}
\eeq
where $\lambda_F=2\pi/k_F$ is the Fermi wavelength and  $\lambda_C=2\pi/mc=2.4\times 10^{-10}$\,cm is the Compton wavelength. For electron number densities in the interval $n=10^{11}-10^{12}$\,cm$^{-2}$, the factor $(\lambda_F/\lambda_C)^2\sim 10^{8}-10^{9}$, and the EDSR signal is stronger than the ESR one by many orders of magnitude, even if SOC is weak.

However, there is a caveat in this estimate, namely, it is valid provided that the Drude part of the conductivity is much smaller then the spin-orbit part,  which imposes rather stringent conditions on the strength SOC and sample quality. Near the resonance, the spin-orbit part of the conductivity can be estimated as\cite{shekhter:2005,maiti:2015}
\bea
\sigma'_{\text{SO}}(\Omega)\sim \frac{e^2}{h} m^*\max\{\alpha^2,\beta^2\}\frac{
\Gamma}{\left(\Omega-\Omega_{\text{r}}\right)^2+
\Gamma^2}.
\eea
Assuming that the linewidth of the resonance is due to D'yakonov-Perel' mechanism in the ballistic regime (cf. Sec.~\ref{sec:damping}), i.e, $\Gamma\sim 1/\tau$, 
and that the Drude conductivity is controlled by the same scattering mechanism, i.e., $\sigma'_{\text{D}}(\Omega_{\text{r}})\sim (e^2/h) E_F/\Omega^2_{\text{r}}\tau$, we obtain for the ratio of the two parts of the conductivity right at the resonance:
\bea
K=\frac{\sigma'_{\text{SO}}(\Omega_{\text{r}})}{\sigma'_{\text{D}}(\Omega_{\text{r}})}\sim 
\frac{m^*\max\{\alpha^2,\beta^2\}}{E_F}
\left(\Omega_{\text{r}}\tau\right)^2.\nn\\
\label{contrast}
\eea
Therefore, even if the resonance is underdamped, i.e., $\Omega_{\text{r}}\tau\gg 1$, it can be still masked by the Drude part if the first factor on the RHS of \Eq{contrast} is sufficiently small. 

Estimates\cite{maiti:2015} show that the resonance in zero magnetic field, when $\Omega_{\text{r}}\sim \max\{\dr,\dd\}$, would be completely masked in a GaAs/GaAlAs heterostructure even with a record-hight mobility of $10^7$\,cm$^2$/V\,s because, due to a relatively weak SOC in this system ($\alpha\sim\beta\sim 1$\,mev\,\AA), $K$ is only $\sim 0.1$.  The problem is further exacerbated by the fact that, in the presence of both Rashba and Dresselhaus SOCs, the lower edge of the continuum is located at $\Omega_{-}=|\alpha-\beta|k_F$, which pushes the energies of the collective modes further down.

A better candidate is an InGaAs/InAlAs quantum well,  where SOC is much stronger, i.e., $\alpha\sim 100\; \mathrm{meV}\,\mathrm{\AA}$  (Ref.~\onlinecite{nitta:1997}), which helps to compensate for smaller mobilities typical for these structures; the highest mobilities reported for InGaAs/InAlAs samples are in the range $\mu=(2-5)\times 10^5 \;\mathrm{cm}^2$/V\,s (Refs.~\onlinecite{sato:2001} and \onlinecite{yamada:2003}). Also, SOC in these structures is predominantly of the Rashba type,\cite{nitta:1997,luo:1990} which alleviates the problem with a competition between the Rashba and Dresselhaus mechanisms.  For a high-mobility InGaAs/InAlAs quantum well, $K\sim 1$ and a detailed calculation confirms that the zero-filed EDSR peak should be visible against the Drude background.\cite{maiti:2015}

\subsubsection{ESR and EDSR in graphene with proximity-induced spin-orbit coupling}
\label{sec:GR_ESR}
\paragraph{Zero magnetic field.}
\begin{figure*}
\centering
\includegraphics[scale=0.5]{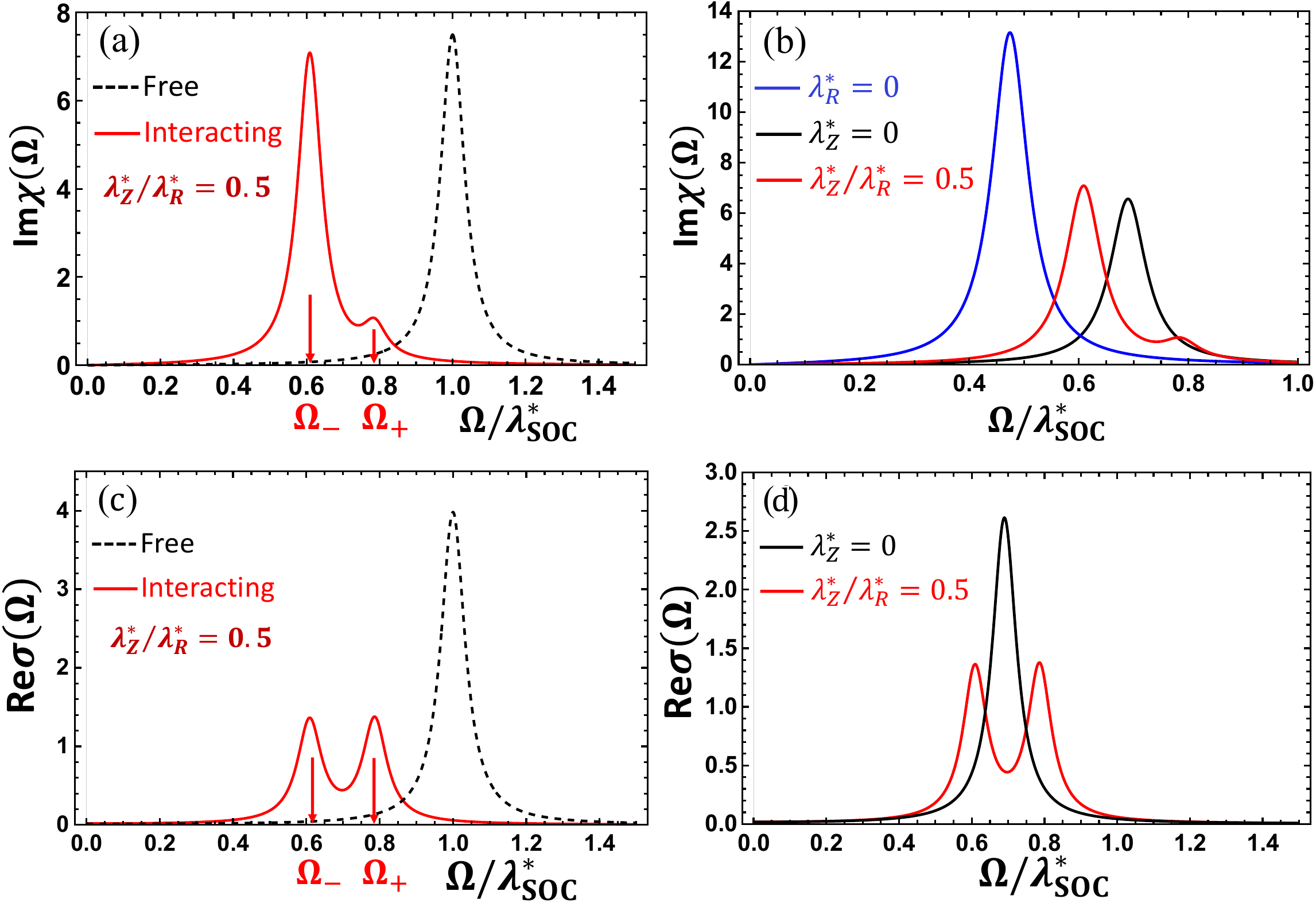}
\caption{\label{Peaks} Theoretical predictions for the zero-field electron spin resonance (ESR) and electric-dipole spin resonance (EDSR) in graphene with proximity-induced spin-orbit coupling (SOC). (a) ESR signal. Vertical axis: the imaginary part of the dynamical spin susceptibility. The frequency on the horizontal axis is 
scaled with $\lambda^*_{\text{SOC}}=\sqrt{\lr^{*2}+\lz^{*2}}$, where $\lr^*$ and $\lz^*$ are (renormalized) couplings of the  Rashba and valley-Zeeman (VZ) types of SOC, respectively. $\Omega_{\pm}$ are the resonance frequencies, given by Eqs.~\er{R+Z} and \er{O0}. Dashed line: non-interacting system. Red solid line: a two-valley Fermi liquid (FL)  with parameters $F^a_0=-0.5500$, $F^a_1=-0.2750$, $F^a_2=-0.1375$,  $H_0=-0.5000$, $H_1=-0.2500$, and $H_2=-0.1250$.  The ratio $\lz^*/\lr^*=0.5$. The choice of FL parameters is the same for all panels of the figure.
 (b) ESR signal in a FL for several values of $\lz^*/\lr^*$, as indicated in the legend. 
(c)  EDSR signal. Vertical axis:  the real part of the optical conductivity. 
Dashed line: non-interacting system. Solid line: FL. 
(d) EDSR signal in a FL for two values of $\lz^*/\lr^*$, as indicated in the legend. To account for smearing of the resonances by disorder,  we added a damping term, $-\delta\hat n(\bk,t)/\tau_s$, to the right-hand side of the kinetic equation \eq{kinetic_edsr}. In all panels of  Fig.~\ref{Peaks}, $1/\tau_s=0.04\lambda^*_{\text{SOC}}$, where $\lambda^*_{\text{SOC}}=\sqrt{\lr^{*2}+\lz^{*2}}$.  For $\lr^*=15.0$\,meV and $\lz^*=7.5$\,meV, the spin relaxation time is $1/\tau_s=1$\,ps. Reprinted with permission from Ref.~\onlinecite{Kumar_TMD:2021}. Copyright 2021 by the American Physical Society.}
\end{figure*}

The zero-field ESR and EDSR in graphene with proximity-induced spin-orbit coupling are predicted to have some interesting features.\cite{Kumar_TMD:2021} In the presence of electron-electron interactions and both types of SOC (Rashba and valley-Zeeman), both ESR and EDSR signals consist of two peaks, centered at the frequencies of coupled oscillations of the uniform and valley-staggered magnetizations, see panels $a$ and $c$ in Fig.~\ref{Peaks}. Splitting of the resonance occurs as long as the Landau function in \Eq{FL} has more than just the $m=0$ harmonic in the spin-exchange and spin-valley-exchange channels, which is always the case for graphene. Next, even if the (interacting) system has only one type of SOC, there are still two resonance modes, $\Omega_{+}$ and $\Omega_{-}$, but one of them is both ESR- and EDSR-silent, because the spectral weights of the corresponding resonances vanish.  This effect is illustrated in panels $b$ and $d$ of Fig.~\ref{Peaks}.  Another interesting feature is that the two ESR (EDSR) peaks have different (comparable) magnitudes. Therefore, EDSR is better way to probe the two-peak structure of the resonance.  

It is worth pointing out that the relative strengths of the Rashba and VZ components of SOC in graphene on TMD substrates is currently an open issue.  While  weak antilocalization experiments on monolayer graphene find VZ SOC to be much stronger than the Rashba one,\cite{wakamura:2018,schonenberger:2018,wakamura:2019} the opposite conclusion is reached in, e.g.,  Refs.~\onlinecite{morpurgo_2015,morpurgo_2016,Yang:2016,Omar:2018}. On the other hand, strong evidence for Rashba SOC being the dominant type in bilayer graphene on WSe$_2$ follows from the dependence of the splitting of the ShdH frequencies on the carrier number density.\cite{morpurgo_2016}
 Without getting deeper into this discussion, we note that the ESR and EDSR experiments can be used as an independent test for the dominant type of SOC. Indeed, the coupling between the electric field and electron spins is possible only due to Rashba SOC. Therefore, if the experiment shows no EDSR signal, while the ESR signal contains only a single peak, this would be a clear indication that VZ SOC is the dominant mechanism.  On the contrary, if single peaks (at the same frequency) are observed both by EDSR and ESR, this would indicate that Rashba SOC is the dominant mechanism. Finally, if both ESR and EDSR signals are split into two peaks, this would indicate that the Rashba and VZ types of SOC are of comparable strength. A quantitative analysis of the signal shape would allow one not only to obtain the spin-orbit coupling constants, but also to extract up to six FL parameters in the $m=0,1,2$ angular momentum channels, which are hard, if at all possible, to be deduced from other types of measurements.
\paragraph{Strong magnetic field.} The opposite case of a strong (compared to SOC) in-plane magnetic field was analyzed in Ref.~\onlinecite{raines:2022} at $q=0$ and at finite $q$ in Ref.~\onlinecite{ullrich:2021}. If the effect of SOC on the spectrum of collective modes is neglected, the latter consists of the Silin modes, corresponding to oscillations of the uniform magnetization with frequencies as in Eq. \er{silin_freq}, and of an additional set of modes, corresponding to oscillations of the valley-staggered magnetization with frequencies
\bea
\tilde\Omega_m=\frac{1+H_m^\perp}{1+F_0^a}\dz,
\eea
where $H_m^\perp$ is the $m^{\text{th}}$ harmonic of the function $H^\perp(\vartheta)$ in \Eq{FL}.
In the absence of SOC, an external magnetic field couples only to the $m=0$ Silin mode while the electric field does not couple to either of the modes. If both Rashba and valley-Zeeman types of SOC are present and, in addition, the Dirac point is gapped due to the breaking of the  A-B symmetry of the honeycomb lattice by the substrate, the external electric field couples to the $m=0$ and $m=2$ Silin modes, and to the $m=0$ and $m=1$ valley-staggered modes.  Therefore, the EDSR spectrum consists, in general, of four peaks. In addition, the resonances occurs not only in the longitudinal conductivity, but also in the transverse (Hall) one, although the external magnetic field does not affect the electron orbits. This last effect occurs due to the Berry curvature of the gapped Dirac point, and its mechanism can be understood already for non-interacting electrons as follows.\cite{raines:2022}

\begin{figure}
    \centering
    \includegraphics[width=\linewidth]{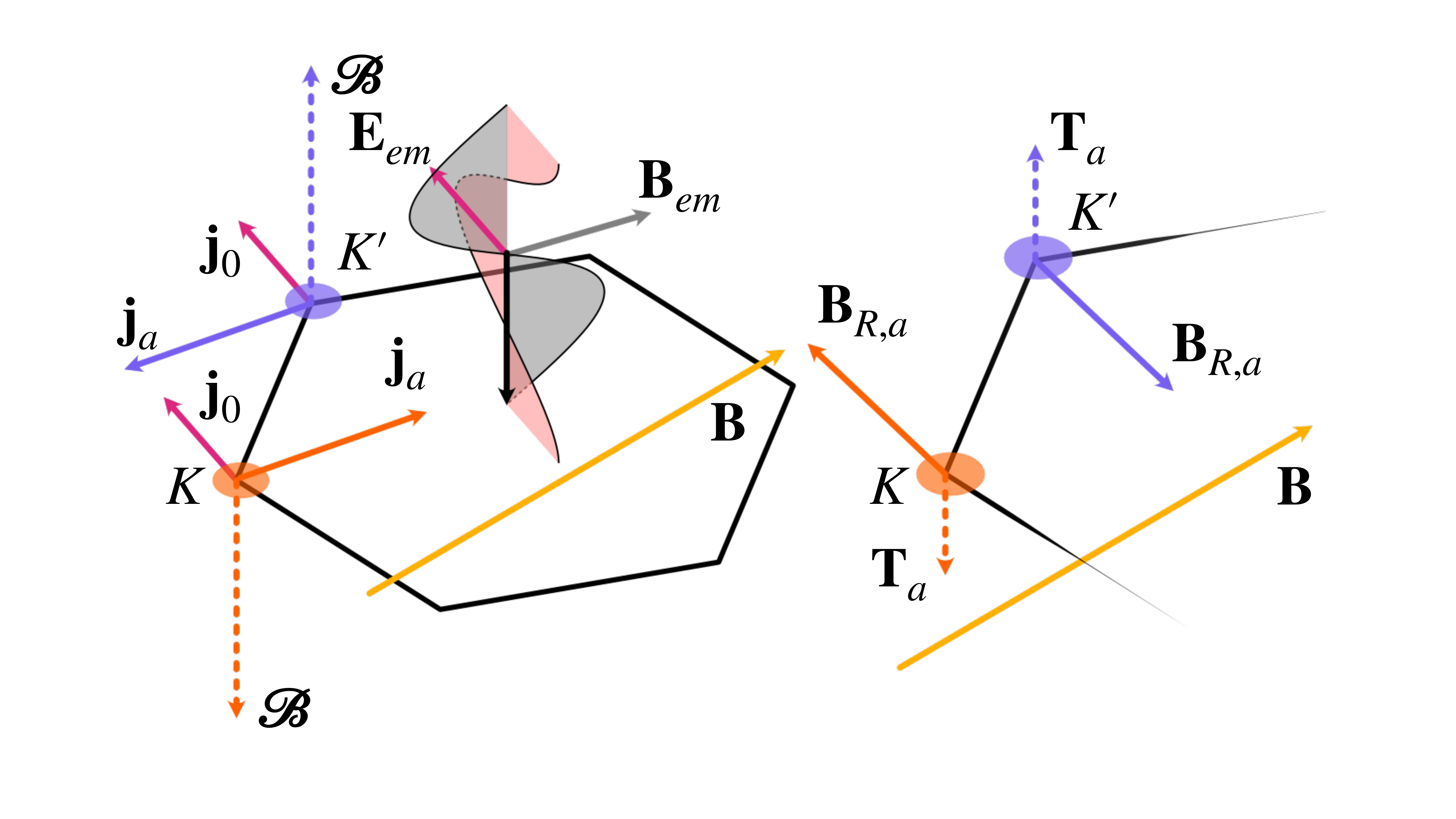}
    \caption{ \emph{Left}: Electric-dipole spin resonance in graphene with proximity-induced spin-orbit coupling in the presence of strong in-plane magnetic field.  The regular, $\mathbf{j}_0 \propto \mathbf{E}_{\text{em}}$, and anomalous, $\mathbf{j}_a \propto \boldsymbol{\mathcal{B}}\times \mathbf{E}_{\text{em}}$, currents at the $K$ (orange) and $K'$ (purple) points in the Brillouin zone, induced by
    the electric field $\mathbf{E}_{\text{em}}$ of the incident electromagnetic wave.
    Here, $\boldsymbol{\mathcal{B}}$ is the valley-staggered Berry curvature.
    \emph{Right}: Anomalous fields and torques.
    Spins are initially polarized along the static magnetic field $\mathbf{B}$.  The anomalous current-induced effective Rashba fields, $\mathbf{B}_{\text{R},a}\propto \lr \left(\hat{\bf z} \times \mathbf{j}_a\right)$, 
    produce
    valley-specific 
    torques $\mathbf{T}_a \propto \mathbf{B} \times \mathbf{B}_{\text{R},a}$, 
    thus exciting the valley-staggered spin mode with an intensity proportional to $|\mathbf{E}_{\text{em}} \times \mathbf{B}|$. Reprinted with permission from Ref.~\onlinecite{raines:2022}. Copyright 2022 by the American Physical Society.
    \label{fig:diagram}}
\end{figure}

Initially, all particle spins are polarized along the external magnetic field, which we take to be along the $\hat{\bf x}$ axis.
Upon application of an external field $\mathbf{E}_{\text{em}}(t)$ the particle spins feel an effective Rashba magnetic field 
$\mathbf{B}_{\text{R}} \propto \hat{\bf z} \times \mathbf{j}$, where $ \mathbf{j}$ is the electric current density, and therefore experience a spin torque
$
 \mathbf{T} \propto \hat{\bf x} \times \mathbf{B}_{\text{R}} \propto \hat{\bf z}  j_x.
 $
The $x$ component of the current $j_x$ is composed of 
regular and anomalous pieces, shown in the left of Fig.~\ref{fig:diagram},
\begin{equation}
j_x(\omega) = -
\frac{e^2 n E_{\text{em},x}}{i\omega m^*} - e^2N{\mathcal B}_z(k)\label{jx}
E_{\text{em},y},
\end{equation}
where $n$ is the number density, $m^*=k_F/\epsilon'_k\vert_{k=k_F}$ is the effective mass, and $\boldsymbol{\mathcal B}(k)=\mp \vd^2\Delta/2(\vd^2k^2+\Delta^2)^{3/2}\hat{\bf z}$ is the Berry curvature of the gapped Dirac points in the $K$ ($K'$) valleys.
The first term in \Eq{jx}  creates identical torques
in both valleys, while the
second one, being proportional to the Berry curvature, yields valley-staggered torques depicted in the right panel of Fig.~\ref{fig:diagram}.
The component of $\mathbf{E}_{\text{em}}$ along $\mathbf{B}$
causes a valley-uniform torque on the spin, exciting the 
Silin mode spin, while
the component of $\mathbf{E}_{\text{em}}$ transverse to $\mathbf{B}$
causes a valley-staggered torque,
and thus excites the valley-staggered spin mode. 
Because the charge-to-spin conversion in both cases is proportional to the Rashba coupling, this leads to a term in the conductivity proportional to $\alpha_R^2$.
Furthermore, the Silin mode contributes to $\sigma_{xx}$, while the valley-staggered mode contributes to $\sigma_{xy}$.

\section{Conclusions}
\label{sec:concl}
In this review, we summarized recent progress in theoretical understanding and experimental observation of a new type of collective spin modes, arising from an interplay between spin-orbit coupling (SOC) and electron-electron interaction. We focused on three types of real systems: i) a two-dimensional (2D) electron gas (2DEG) with Rashba and/or Dresselhaus SOC, ii) graphene with proximity-induced SOC, and iii) the Dirac state on the surface of a three-dimensional topological insulator. Provided that SOC and/or external magnetic field are weak, i.e., the corresponding energy scales are much smaller than the Fermi energy, collective modes in systems i) and ii) can be described within the single-valley or two-valley versions of the Fermi-liquid (FL) theory, respectively.  A transparent physical picture of such collective modes arises due to mapping of a kinetic equation for a 2D FL onto an effective tight-binding model for an artificial one-dimensional lattice, whose sites are labeled by the projections of angular momentum on the normal to the 2DEG plane ($m$). Rashba SOC plays the role of on-site energies, while Zeeman and Dresselhaus terms correspond to hopping between the nearest and next-to-nearest neighbors, respectively, whereas the $m$-dependent components of the Landau interaction function create ``defects'' of both on-site and bond types. Within this mapping, the continuum of particle-hole excitations plays a role of the conduction band, while collective modes are the bound states produced by defects. 

We discussed the results of recent Raman experiments on Cd$_{1-x}$Mn$_x$Te quantum wells\cite{perez:2007, baboux:2013,baboux:2015,perez:2016,perez:2017,perez:2019} and the Dirac state on the surface of Bi$_2$Se$_3$,\cite{kung:2017} in which some of the predicted collective modes have been observed, and formulated predictions for future electron spin resonance (ESR) and electric-dipole spin resonance (EDSR) experiments on graphene with proximity-induced SOC.

The new type of collective  modes, discussed in this paper, may have potential applications in spintronics, magnonics, optoelectronics, and quantum sensing. Indeed, such modes can be thought of as massive ``particles'', with masses fixed by the FL interaction, moving in a potential profile produced by SOC.\cite{ashrafi:2012} By modulating the strength of SOC along the plane of motion, e.g., by gating, one can confine the modes to waveguides and use them to transmit signals. Despite inherent disorder and other sources of damping, the so far observed modes of this type are quite sharp and robust; for example, the collective mode on the surface of Bi$_2$Se$_3$ is observed up to 300\,K.\cite{kung:2017} 


\acknowledgments
It is our great pleasure and honor to dedicate this paper to the 95th birthday of Professor Emmanuel Iosifovich Rashba, whose seminal work on spin-orbit effects in solids laid the foundation of modern spintronics. 
One of us (D.L.M.) has had a pleasure of knowing Emmanuel Iosifovich for a long time and have been fortunate to collaborate with him on Ref.~\onlinecite{ashrafi:2013}, which was one of the most gratifying collaborations in his (D.L.M.'s) career. Emmanuel Iosifovich impresses anyone whom he interacts with even briefly by the clarity of his thoughts, encyclopedic knowledge of physics, acute attention to details, and outward friendliness.

We are grateful to our co-authors A. Ashrafi (Magine), G. Blumberg, L. Glazman, A. Goyal, S. Kung, Z. Raines, E. Rashba, and P. Sharma for collaborating with us on the projects discussed in this paper, and to  H. Bouchiat, C. Bowers, A. Chubukov, S. Gu{\'e}ron, F. Perez, C. Ullrich,  
E. Sherman, O. Starykh, 
and T. Wakamura for stimulating discussions. This work was supported by the National Science Foundation under Grant No. NSF DMR-1720816 (D.L.M.), the Department of Energy (DOE) Basic Energy Sciences under Grant No. DE-SC0020353 (A.K.), and the Natural Sciences and Engineering Research Council of Canada (NSERC) Grant No. RGPIN-2019-05486 (S.M.).

\newpage
\bibliography{dm_JETP,referenceFile,Ashrafi_SOFL,ref_2015}
\end{document}